\documentclass[12pt]{article}
\usepackage{putex}
\usepackage{graphicx}
\begin{document}

\preprint{hep-th/0607022 \\ PUPT-2201}

\institution{PU}{Joseph Henry Laboratories, Princeton University, Princeton, NJ 08544}

\title{The stress tensor of a quark moving through ${\cal N}=4$ thermal plasma}

\authors{Joshua J. Friess, Steven S. Gubser, Georgios Michalogiorgakis, and \\[10pt] Silviu S. Pufu}

\abstract{\noindent We develop the linear equations that describe graviton perturbations of $AdS_5$-Schwarzschild generated by a string trailing behind an external quark moving with constant velocity.  Solving these equations allows us to evaluate the stress tensor in the boundary gauge theory.  Components of the stress tensor exhibit directional structures in Fourier space at both large and small momentum.  We comment on the possible relevance of our results to relativistic heavy ion collisions.}


\maketitle
\tableofcontents

\section{Introduction}
\label{INTRODUCTION}

If an external quark (meaning an infinitely massive, fundamentally charged, point-like particle) is passed through a thermal plasma of $SU(N)$ ${\cal N}=4$ super-Yang-Mills with both $N$ and $g_{YM}^2 N$ large, it experiences a drag force \cite{Herzog:2006gh,Gubser:2006bz}
 \eqn{DragForce}{
  {dp \over dt} =
    -{\pi \sqrt{g_{YM}^2 N} \over 2} T^2 {v \over \sqrt{1-v^2}} \,,
 }
where $v$ is the speed of the quark and $T$ is the temperature of the plasma.  The result \eno{DragForce} was derived using the duality between string theory in anti-de Sitter space and conformal field theory on the boundary of anti-de Sitter space (AdS/CFT) \cite{Maldacena:1997re,Gubser:1998bc,Witten:1998qj}.  The diffusion constant $D = 2/(\pi T \sqrt{g_{YM}^2 N})$ implied by \eno{DragForce} was derived independently in \cite{Casalderrey-Solana:2006rq}, also using AdS/CFT\@.  A number of other papers \cite{Sin:2004yx, Liu:2006ug,Buchel:2006bv,Herzog:2006se,Caceres:2006dj,Friess:2006aw,Sin:2006yz,Gao:2006se, Armesto:2006zv, Peeters:2006iu, Avramis:2006ip, Lin:2006au, Caceres:2006as, Vazquez-Poritz:2006ba} use AdS/CFT in related ways to describe energy dissipation from a moving quark.  This work is motivated in part by the phenomenon of jet-quenching observed in RHIC experiments: for recent experimental accounts see for example \cite{Adler:2002tq,Wang:2004kf,Adler:2005ee}.

It was observed in \cite{Herzog:2006gh,Gubser:2006bz} that $\langle T_{mn} \rangle$ in the boundary gauge theory is calculable via AdS/CFT but requires a technically non-trivial analysis of graviton perturbations in $AdS_5$-Schwarzschild.  The aim of this paper is to develop the relevant equations and solve them, both in limits that are analytically tractable and through use of numerics for selected values of the velocity.  Section~\ref{EOMS} comprises a derivation of the equations.  The linearized graviton equations are stated in full at the end of section~\ref{EOMS}.  Section~\ref{ASYMPTOTICS} includes solutions of the equations near the boundary and near the horizon of $AdS_5$-Schwarzschild, as well as expressions for the stress tensor in the near-field limit.  A boundary-value problem in classical five-dimensional gravity is stated at the end of section~\ref{HORIZON} which determines $\langle T_{mn} \rangle$ in the boundary theory.  Section~\ref{NUMERICS} presents the results of numerical work, and section~\ref{DISCUSSION} is devoted to discussion of the possible relevance to jet-quenching in relativistic heavy-ion collisions.

The reader wishing to skip all the five-dimensional technicalities and see the ``answers'' may skip directly to section~\ref{NUMERICS} with the following definitions and conventions in mind:
 \begin{itemize}
  \item Wave-numbers $\vec{K} = (K_1,K_2,K_3)$ are rendered dimensionless by including a factor of $z_H = 1/\pi T$.
  \item Usually we set $K_3=0$ and $K_2 = K_\perp > 0$.
  \item Often we refer to $K = \sqrt{K_1^2 + K_\perp^2}$ and $\theta = \tan^{-1} (K_\perp/K_1)$.
  \item $\langle T^K_{mn} \rangle$ is the $K$-th Fourier coefficient of the co-moving part of $\langle T_{mn} \rangle$ in the $3+1$-dimensional boundary gauge theory: see \eno{FullTmn}.
  \item $Q^{\rm tot}_{mn}$ is a dimensionless quantity proportional to $\langle T^K_{mn} \rangle$: see \eno{GotTmn}.
  \item Using symmetries and conservation laws, the non-zero components of $Q^{\rm tot}_{mn}$ can be expressed in terms of three complex quantities $Q^{\rm tot}_A$, $Q^{\rm tot}_D$, and $Q^{\rm tot}_E$: see \eno{QtoQ}.
  \item $Q^{\rm tot}_E$ is the easiest quantity to interpret, as it is directly proportional to $\langle T^K_{00} \rangle$, the $K$-th Fourier component of the energy density.
  \item $Q^K_X$ for $X = A$, $D$, or $E$ is $Q^{\rm tot}_X$ with the Coulombic near-field of the quark subtracted away: see \eno{QXdef}.  Note also that the inhomogeneous term $p_{mn}$ in \eno{QtoQ} amounts to far-field subtraction in the definition of both $Q^{\rm tot}_X$ and $Q^K_X$.
 \end{itemize}

\section{The graviton equations of motion}
\label{EOMS}

The relevant part of the action for supergravity plus the string is
 \eqn{GravAction}{
  S = \int d^5 x \left[ {\sqrt{-G} \, (R+12/L^2) \over 2\kappa_5^2} -
   {1 \over 2\pi\alpha'} \int d^2\sigma \, \sqrt{-g} \,
    \delta^5(x^\mu-X^\mu(\sigma))
  \right] \,,
 }
where $L$ is the radius of $AdS_5$.  We have excluded the dilaton from \eno{GravAction} because it decouples from the metric at the level of linear perturbations around the $AdS_5$-Schwarzschild background,
 \eqn{BackgroundMetric}{
  ds_{(0)}^2 = G^{(0)}_{\mu\nu} dx^\mu dx^\nu
   = {L^2 \over z_H^2 y^2} \left( -h dt^2 + d\vec{x}^2 +
     z_H^2 {dy^2 \over h} \right) \qquad h \equiv 1-y^4 \,.
 }
We have introduced in \eno{BackgroundMetric} a radial variable $y$ which runs from $0$ at the boundary of $AdS_5$-Schwarzschild to $1$ at the horizon.  A more conventional choice of radial variable is $z = z_H y$.  The position of the string can be described in static gauge as
 \eqn[c]{Xstatic}{
  X^\mu(t,y) \equiv
   \begin{pmatrix} t & X^1(t,y) & 0 & 0 &
     y \end{pmatrix}  \cr
  X^1(t,y) = vt + \xi(y) \qquad
  \xi(y) = -{z_H v \over 4i} \left( \log {1-iy \over 1+iy} +
    i \log {1+y \over 1-y} \right) \,.
 }
The equation of motion is
 \eqn{EinsteinFive}{
  R^{\mu\nu} - {1 \over 2} G^{\mu\nu} R - {6 \over L^2}
    G^{\mu\nu} = \tau^{\mu\nu} \,,
 }
where
 \eqn{taumn}{
  \tau^{\mu\nu} &= -{\kappa_5^2 \over 2\pi\alpha'}
     \int d^2 \sigma \, {\sqrt{-g} \over \sqrt{-G}} \,
      \delta^5(x^\mu-X^\mu) \partial_\alpha X^\mu
       \partial^\alpha X^\nu  \cr
  &= -{\kappa_5^2 \over 2\pi\alpha'} {z_H^3 \over L^3} y^3
   \sqrt{1-v^2} \delta(x^1-vt-\xi(y)) \delta(x^2) \delta(x^3)
     \partial_\alpha X^\mu \partial^\alpha X^\nu \,,
 }
is the stress-energy tensor due to the string, expressed explicitly in static gauge in the second line.  If we perturb
 \eqn{Gperturb}{
  G_{\mu\nu} = G_{\mu\nu}^{(0)} + h_{\mu\nu} \,,
 }
then, schematically, the form of the linearized equations following from \eno{EinsteinFive} is
 \eqn{LinEOM}{
  \Delta_{AdS} h^{\mu\nu} = \tau^{\mu\nu}
 }
where $h^{\mu\nu} = G^{\mu\rho}_{(0)} G^{\nu\sigma}_{(0)} h_{\rho\sigma}$ and $\Delta_{AdS}$ is a variant of the Lichnerowicz operator.

The stress tensor $\tau^{\mu\nu}$ depends on $x^1$ and $t$ only through the combination $x^1-vt$.  Thus we can expand
 \eqn{FourierTau}{
  \tau^{\mu\nu}(t,x^1,x^2,x^3,y) = \int {d^3 K \over (2\pi)^3} \,
    \tau_K^{\mu\nu}(y) \, e^{i \left[ K_1 (x^1-vt) +
      K_2 x^2 + K_3 x^3 \right] / z_H} \,.
 }
(Note that $\vec{K} = z_H \vec{k} = \vec{k}/\pi T$ is dimensionless.)  If our interest is the co-moving graviton response, we can make a similar expansion for $h_{\mu\nu}$.  Then, again schematically, one obtains from \eno{LinEOM} a set of coupled ordinary differential equations in $y$:
 \eqn{LinODEs}{
  {\cal E}^{\mu\nu} \equiv
    \Delta_{AdS}^K h^{\mu\nu}_K - \tau^{\mu\nu}_K = 0 \,.
 }
It is these equations which we wish to formulate more precisely and then solve.  Note that although the Fourier modes $\tau_K^{\mu\nu}$ and $h_K^{\mu\nu}$ are complex, they satisfy conditions like $\tau^{\mu\nu}_{-K} = (\tau^{\mu\nu}_K)^*$ that ensure the position space quantities are real.  From the asymptotic behavior of $h_K^{\mu\nu}$ near the boundary of $AdS_5$-Schwarzschild one may extract the $K$-th Fourier mode $\langle T^K_{mn} \rangle$ of the co-moving contribution to the stress tensor.  A detailed discussion of the extraction of $\langle T^K_{mn} \rangle$ is deferred to section~\ref{BOUNDARY}.

The rotational symmetry around the axis of motion of the quark enables us to choose $\vec{K} = (K_1,K_\perp,0)$ with $K_\perp \geq 0$.  The remaining symmetry is a ${\bf Z}_2$ sending $x^3 \to -x^3$.  The metric perturbation can be parametrized as
 \eqn{Hparam}{
  h^K_{\mu\nu} = {\kappa_5^2 \over 2\pi\alpha'}
    {1 \over \sqrt{1-v^2}} {L \over z_H^2 y^2}
   \begin{pmatrix} H_{00} & H_{01} & H_{02} & H_{03} & 0  \\
     H_{10} & H_{11} & H_{12} & H_{13} & 0  \\
     H_{20} & H_{21} & H_{22} & H_{23} & 0  \\
     H_{30} & H_{31} & H_{32} & H_{33} & 0  \\
     0 & 0 & 0 & 0 & 0 \end{pmatrix} \,.
 }
The vanishing entries represent a gauge choice which we will refer to as axial gauge.  The $K$-th Fourier mode of the string's stress tensor is
 \eqn{StringStress}{
  \tau_K^{\mu\nu} = {\kappa_5^2 \over 2\pi\alpha'}
   {e^{-i K_1 \xi(y)/z_H} \over \sqrt{1-v^2}} {y^5 \over L^5}
   \begin{pmatrix} z_H^2 \displaystyle{h + v^2 y^4 \over h^2} &
      z_H^2 \displaystyle{v \over h} & 0 & 0 &
      z_H \displaystyle{v^2 y^2 \over h}  \\[5\jot]
     z_H^2 \displaystyle{v \over h} & z_H^2 v^2 & 0 & 0 & z_H v y^2
       \\[3\jot]
     0 & 0 & 0 & 0 & 0  \\[2\jot]
     0 & 0 & 0 & 0 & 0  \\[3\jot]
     z_H \displaystyle{v^2 y^2 \over h} &
       z_H v y^2 & 0 & 0 & v^2 - h
   \end{pmatrix} \,,
 }
where $\xi(y)$ is as given in \eno{Xstatic}.

Because \eno{LinODEs} is an equation for symmetric rank-two tensors, it has $15$ algebraically independent component equations.  Ten of these, namely ${\cal E}^{mn} = 0$ with $0 \leq m \leq n \leq 3$, are second order, and the other five, ${\cal E}^{\mu 5}=0$ with $\mu$ unrestricted, are first order constraints.\footnote{We associate $y$ with $\mu=5$.  There is no coordinate associated with $\mu=4$.  This convention serves as a reminder that $y$ is the fifth dimension.}  There are $10$ dependent variables $H_{mn}$, so the full system ${\cal E}^{\mu\nu} = 0$ seems overdetermined.  But it isn't: if the constraints are imposed at one value of $y$ and the second order equations are then solved, the constraints continue to hold automatically for all $y$.

The ${\bf Z}_2$ symmetry that takes $x^3 \to -x^3$ causes the equations \eno{LinODEs} to partially decouple.  The ${\bf Z}_2$ ``charge'' of a component $H_{mn}$ of the metric perturbation is the parity of the number of indices equal to $3$.  Similar charge assignments can be made to the equations ${\cal E}^{\mu\nu}=0$.  Thus for example $H_{03}$ and ${\cal E}^{13}=0$ are odd while $H_{33}$ and ${\cal E}^{01}=0$ are even.  The three odd second order equations of motion and the one odd constraint equation involve only the odd variables, whereas the seven even second order equations and the four even constraint equations involve only the even variables.  Moreover, only the even equations involve non-zero components of the stress tensor \eno{StringStress}.  So it is consistent to set all the odd variables equal to $0$ from the outset.  In the interests of generality, we will not do this yet, but rather consider how the equations may be further decoupled.

Briefly, we make the following definitions and find the following differential equations:
 \eqn{KTdef}{
  K = \sqrt{K_1^2 + K_\perp^2} \qquad
  \theta = \tan^{-1} {K_\perp \over K_1}
 }
 \eqn{AAdef}{
  A = {-H_{11} + 2 \cot\theta H_{12} -
    \cot^2\theta H_{22} + \csc^2\theta H_{33} \over 2v^2}
 }
 \eqn{AAeom}{
  \left[ \partial_y^2 + \left( -{3 \over y} + {h' \over h}
    \right) \partial_y + {K^2 \over h^2}
     (v^2 \cos^2\theta - h) \right] A =
      {y \over h} e^{-i K_1 \xi/z_H}
 }
 \eqn{BBdef}{
  B_1 = {H_{03} \over K^2 v} \qquad
  B_2 = -{H_{13} + \tan\theta H_{23} \over K^2 v^2}
 }
 \eqn{BBeom}{
  \left[ \partial_y^2 +
   \begin{pmatrix} -{3 \over y} & 0 \\
    0 & -{3 \over y} + {h' \over h} \end{pmatrix} \partial_y +
   {K^2 \over h^2} \begin{pmatrix} -h & v^2 \cos^2\theta h  \\
     -1 & v^2 \cos^2\theta \end{pmatrix} \right]
   \begin{pmatrix} B_1 \\ B_2 \end{pmatrix} =
   \begin{pmatrix} 0 \\ 0 \end{pmatrix}
 }
 \eqn{BBconstraint}{
  B_1' - h B_2' = 0
 }
 \eqn{CCdef}{
  C = {-\sin\theta H_{13} + \cos\theta H_{23}  \over K}
 }
 \eqn{CCeom}{
  \left[ \partial_y^2 + \left( -{3 \over y} + {h' \over h}
    \right) + {K^2 \over h^2} (v^2 \cos^2\theta - h) \right] C = 0
 }
 \eqn{DDdef}{
  D_1 = {H_{01} - \cot\theta H_{02} \over 2v} \qquad
  D_2 = {-H_{11} + 2 \cot2\theta H_{12} + H_{22} \over 2v^2}
 }
 \eqn{DDeom}{
  \left[ \partial_y^2 +
   \begin{pmatrix} -{3 \over y} & 0 \\
    0 & -{3 \over y} + {h' \over h} \end{pmatrix} \partial_y +
   {K^2 \over h^2} \begin{pmatrix} -h & v^2 \cos^2\theta h  \\
     -1 & v^2 \cos^2\theta \end{pmatrix} \right]
   \begin{pmatrix} D_1 \\ D_2 \end{pmatrix} =
   {y \over h} e^{-i K_1 \xi/z_H}
    \begin{pmatrix} 1 \\ 1 \end{pmatrix}
 }
 \eqn{DDconstraint}{
  D_1' - h D_2' = {y^3 \over i vK_1} e^{-i K_1 \xi/z_H}
 }
 \eqn[c]{EEdef}{
  E_1 = {1 \over 2} \left( -{3 \over h} H_{00} + H_{11} +
    H_{22} + H_{33} \right) \qquad
  E_2 = {H_{01} + \tan\theta H_{02} \over 2v}  \cr
  E_3 = {H_{11} + H_{22} + H_{33} \over 2}  \qquad
  E_4 = {-H_{11} - H_{22} + 3\cos2\theta (-H_{11} + H_{22}) +
    2 H_{33} - 6 \sin2\theta H_{12} \over 4}
 }
 \eqn{EEeom}{
  &\left[ \partial_y^2 +
   \begin{pmatrix} -{3 \over y} + {3h' \over 2h} & 0 & 0 & 0  \\
     0 & -{3 \over y} & 0 & 0  \\
     0 & 0 & -{3 \over y} + {h' \over 2h} & 0  \\
     0 & 0 & 0 & -{3 \over y} + {h' \over h} \end{pmatrix}
      \partial_y \right.  \cr
  &\qquad\qquad\left. {} +
   {K^2 \over 3h^2} \begin{pmatrix}
    -2h & 12 v^2 \cos^2\theta & 6v^2 \cos^2\theta + 2 h & 0  \\
    0 & 0 & 2h & h  \\
    0 & 0 & -2h & -h  \\
    2h & -12 v^2 \cos^2\theta & 0 & 3 v^2 \cos^2\theta + h
   \end{pmatrix} \right]
  \begin{pmatrix} E_1 \\ E_2 \\ E_3 \\ E_4 \end{pmatrix}  \cr
  &\qquad\qquad\qquad\qquad =
   {y \over h} e^{-i K_1 \xi/z_H}
   \begin{pmatrix} 1 + {v^2 \over h}  \\
    1 \\ -1+v^2 - {v^2 \over h}  \\
    v^2 {1+3\cos2\theta \over 2} \end{pmatrix}
 }
 \eqn{EEconstraint}{
  &\left[ \begin{pmatrix} 0 & 1 & 1 & 0 \\
   -h & 0 & -3v^2 \cos^2\theta - h & -h \\
   h & 0 & 2 & 0 \end{pmatrix} \partial_y \right.
    \cr
  &\qquad\qquad \left. {} +
   {1 \over 6h} \begin{pmatrix}
    0 & -6h' & -3h' & 0  \\
    -3 hh' & 18v^2 \cos^2\theta h' & 3 (3v^2 \cos^2\theta + h) h'
     & 0  \\
    2 K^2 yh & -12 K^2 v^2 y \cos^2\theta &
     -2 K^2 y (3v^2 \cos^2\theta - h) & 2 K^2 y h
   \end{pmatrix} \right]
   \begin{pmatrix} E_1 \\ E_2 \\ E_3 \\ E_4 \end{pmatrix}  \cr
  &\qquad\qquad\qquad\qquad =
   {h' \over 4Kyh} e^{-i K_1 \xi/z_H}
   \begin{pmatrix} -ivy \sec\theta  \\
    3 ivy \cos\theta (v^2+h)  \\
    K(v^2 - h) \end{pmatrix} \,.
 }
Let us summarize the salient features of these equations:
 \begin{itemize}
  \item The $15$ equations ${\cal E}^{\mu\nu} = 0$ split up into five sets, decoupled from one another.
  \item The $B$ and $C$ sets \eno{BBdef}-\eno{CCeom} involve only the ${\bf Z}_2$-odd components of the metric, and so it is inevitable that they are homogeneous.  We may set $B_1=B_2=C=0$ and focus on the $A$, $D$, and $E$ equations.
  \item The $A$ equation \eno{AAeom} happens to be identical to the dilaton equation of motion up to a factor multiplying the source term, so we may borrow directly from \cite{Friess:2006aw} to find its solution.  The $C$ equation \eno{CCeom} is also the same as the dilaton equation except that it is homogeneous.
  \item The $B$ and $D$ equations, \eno{BBeom}-\eno{BBconstraint} and \eno{DDeom}-\eno{DDconstraint}, are identical except that the former are homogeneous and the latter are not.  Each set involves one constraint and two second order equations of motion.
  \item The $E$ set \eno{EEdef}-\eno{EEconstraint} involves four second order equations of motion and three constraints.
  \item The total momentum $K$ enters the equations of motion only as a multiplicative factor on the non-derivative coefficient matrices and through the source terms.  Elaborations of the WKB method may therefore be suitable for approximately solving the equations at large $K$, and series solutions in $K$ may be used at small $K$.  Section~\ref{SMALLK} includes further discussion of small $K$ approximations.
 \end{itemize}

\section{Analytic approximations}
\label{ASYMPTOTICS}

Although the discussion at the end of the previous section allows us to set the three odd $H_{mn}$ to $0$, and correspondingly $B_1=B_2=C=0$, we sometimes refrain from doing so in the following discussion of limiting forms of the equations.

\subsection{Near the boundary}
\label{BOUNDARY}

By solving the equations of motion in a series expansion in $y$ one obtains the leading forms
 \eqn{XsetBoundary}{
  X &= -{P_X \over 3} y^3 + Q^{\rm tot}_X y^4 + R_X \qquad
    X = A,B_1,B_2,C,D_1,D_2,E_1,E_2,E_3,E_4 \,,
 }
where
 \eqn[c]{Pvalues}{
  P_A = P_{D_1} = P_{D_2} = P_{E_2} = -P_{E_3} = 1 \qquad
   P_{E_1} = 1+v^2 \qquad P_{E_4} = v^2 (3\cos^2\theta - 1)  \cr
  P_{B_1} = P_{B_2} = P_C = 0 \,.
 }
The $Q^{\rm tot}_X$ are integration constants related to the VEV's of the stress tensor.  The $R_X$ are integration constants which can be set to zero because non-zero values would correspond to deformations of the gauge theory lagrangian.  The constraint equations imply relations among the $Q^{\rm tot}_X$:
 \eqn[c]{ConstraintsOnQ}{
  Q^{\rm tot}_{D_1}-Q^{\rm tot}_{D_2} = {\sec\theta \over 4 i v K}
    \qquad
  Q^{\rm tot}_{B_1}-Q^{\rm tot}_{B_2} = 0  \cr\noalign{\vskip2\jot}
  Q^{\rm tot}_{E_1} - 2 Q^{\rm tot}_{E_2} = {v \sec\theta \over 2 i K}
   \qquad
  Q^{\rm tot}_{E_1} + 2 Q^{\rm tot}_{E_3} = 0  \cr
  (1-3v^2 \cos^2\theta) Q^{\rm tot}_{E_1} + 2 Q^{\rm tot}_{E_4} =
    {3iv(1+v^2) \cos\theta \over 2K} \,.
 }
The meaning of the equations \eno{ConstraintsOnQ} becomes clearer in terms of the original variables $H_{mn}$, whose series expansion near the boundary of $AdS_5$-Schwarzschild includes the leading terms
 \eqn{BdySolns}{
  H_{mn} = -{P_{mn} \over 3} y^3 + Q^{\rm tot}_{mn} y^4 +
   R_{mn} \,.
 }
$P_{mn}$, $Q^{\rm tot}_{mn}$, and $R_{mn}$ are linear combinations, respectively, of $P_X$, $Q^{\rm tot}_X$, and $R_X$, as can be deduced by inverting the relations \eno{AAdef}, \eno{BBdef}, \eno{CCdef}, \eno{DDdef}, and \eno{EEdef}.  In particular, after setting ${\bf Z}_2$-odd quantities to zero and using \eno{ConstraintsOnQ} to eliminate $Q^{\rm tot}_{D_2}$, $Q^{\rm tot}_{E_2}$, $Q^{\rm tot}_{E_3}$, and $Q^{\rm tot}_{E_4}$ in favor of $Q^{\rm tot}_A$, $Q^{\rm tot}_{D_1} \equiv Q^{\rm tot}_D$, and $Q^{\rm tot}_{E_1} \equiv Q^{\rm tot}_E$, one obtains
 \eqn{QtoQ}{
  Q^{\rm tot}_{mn} = a_{mn} Q^{\rm tot}_A +
   d_{mn} Q^{\rm tot}_D + e_{mn} Q^{\rm tot}_E + p_{mn}
 }
where
 \eqn{amnDef}{
  \begin{pmatrix} a_{mn} \end{pmatrix} =
   {v^2 \sin^2\theta \over 2} \begin{pmatrix}
     0 & 0 & 0  \cr
     0 & -2 \sin^2\theta & \sin 2\theta & 0  \cr
     0 & \sin 2\theta & -2 \cos^2\theta & 0  \cr
     0 & 0 & 0 & 2 \end{pmatrix}
 }
 \eqn{dmnDef}{
  \begin{pmatrix} d_{mn} \end{pmatrix} =
   {v \over 2} \begin{pmatrix}
     0 & 4\sin^2\theta & -2\sin 2\theta & 0  \cr
     4\sin^2\theta & -2v \sin^2 2\theta & v \sin 4\theta & 0  \cr
     -2 \sin 2\theta & v \sin 4\theta & 2v \sin^2 2\theta & 0  \cr
     0 & 0 & 0 & 0 \end{pmatrix}
 }
{\small
 \eqn[c]{emnDef}{
  \begin{pmatrix} e_{mn} \end{pmatrix} =
   {1 \over 4} \begin{pmatrix}
     -4 & 4v \cos^2 \theta & 2v \sin 2\theta & 0  \cr
     4v \cos^2 \theta & 4 e_{11} &
       (1-3v^2 \cos^2 \theta) \sin 2\theta & 0  \cr
     2v \sin 2\theta & (1-3v^2 \cos^2 \theta) \sin 2\theta &
      4 e_{22} & 0  \cr
     0 & 0 & 0 & -2 + 2v^2 \cos^2 \theta \end{pmatrix}  \cr
  e_{11} = {1 \over 2} \left[ -1 + (1+v^2) \cos^2 \theta -
       3v^2 \cos^4 \theta \right]  \cr
  e_{22} = {1 \over 2} \cos^2 \theta (-1-2v^2 + 3v^2 \cos^2 \theta)
 }
 \eqn{pmnDef}{
  \begin{pmatrix} p_{mn} \end{pmatrix} =
   {iv \cos\theta \over 4K} \begin{pmatrix}
    0 & 2v & 2v \tan\theta & 0  \cr
    2v & -3+v^2 + (1-3v^2) \cos^2 \theta &
     \left[ -2 + (1-3v^2) \cos^2 \theta \right] \tan\theta & 0  \cr
    2v \tan\theta &
     \left[ -2 + (1-3v^2) \cos^2 \theta \right] \tan\theta &
     2-2v^2 - (1-3v^2) \cos^2 \theta & 0  \cr
    0 & 0 & 0 & 1+v^2 \end{pmatrix}
 }
}
The $Q^{\rm tot}_{mn}$ are integration constants which are proportional to entries of $\langle T^K_{mn} \rangle$:
 \eqn{GotTmn}{
  \langle T^K_{mn} \rangle = {\pi^3 T^4 \sqrt{g_{YM}^2 N} \over
   \sqrt{1-v^2}} Q^{\rm tot}_{mn} \,.
 }
What we mean by $\langle T^K_{mn} \rangle$ is a co-moving Fourier coefficient of the quark's contribution to $\langle T_{mn} \rangle$.  The overall factor in \eno{GotTmn} can be determined through first principles along the lines of \cite{Balasubramanian:1999re}, but we find it more instructive to obtain it heuristically by considering what may seem at first to be a digression: $AdS_5$-Schwarzschild in axial gauge.

Defining a new radial variable $q$ through
 \eqn{AxialQ}{
  y^2 = {q^2 \over 1+q^4/4} \,,
 }
one finds that the line element \eno{BackgroundMetric} becomes
 \eqn{AdSSchAxial}{
  ds^2 &= {L^2 \over z_H^2 q^2} (-dt^2 + d\vec{x}^2 + z_H^2 dq^2) +
    h_{\mu\nu} dx^\mu dx^\nu \cr
  h_{\mu\nu} &= {L^2 q^2 \over 4 z_H^2} \diag\left\{
    {3-q^4/4 \over 1+q^4/4},1,1,1,0 \right\} \,.
 }
This is indeed an axial gauge description of $AdS_5$-Schwarzschild because $h_{\mu 0} = 0$; note however that \eno{AdSSchAxial} is an exact rewriting of \eno{BackgroundMetric}.

On general grounds, the stress tensor of the boundary theory must be proportional to the coefficient of $q^4$ in $h_{mn}$.  But in the case of $AdS_5$-Schwarzschild, there is a pre-existing determination of the energy density and pressure based on \cite{Gubser:1996de}:
 \eqn{DetermineEP}{
  {\epsilon \over 3} = p = {\pi^2 \over 8} N^2 T^4 \,.
 }
Therefore we conclude that
 \eqn{DeterminedTmn}{
  \langle T_{mn} \rangle = {\pi^2 \over 8} N^2 T^4
    \lim_{q \to 0} {z_H^2 \over q^3 L^2} \partial_q (q^2 h_{mn}) \,.
 }
Returning to the setup with a string dangling into $AdS_5$-Schwarzschild means that on top of the ``perturbation'' $h_{\mu\nu}$ that deforms $AdS_5$ into $AdS_5$-Schwarzschild we must add an additional perturbation, namely the $h_{\mu\nu}$ whose Fourier coefficients are given in \eno{Hparam}.  The result \eno{DeterminedTmn} applies unchanged, except that the limit exists only after certain divergent delta-function contributions have been subtracted.  After using \eno{BdySolns} and the standard relations
 \eqn{BasicRelations}{
  N^2 \kappa_5^2 = 4\pi^2 L^3 \qquad
   {L^2 \over \alpha'} = \sqrt{g_{YM}^2 N}
 }
one obtains
 \eqn{FullTmn}{
  \langle T_{mn} \rangle = {\pi^2 \over 8} N^2 T^4
    \diag\{ 3,1,1,1 \} +
   \int {d^3 K \over (2\pi)^3} \langle T^K_{mn} \rangle
     e^{i \left[ K_1 (x^1-vt) + K_2 x^2 + K_3 x^3 \right]/z_H}
 }
where $\langle T^K_{mn} \rangle$ is indeed given by \eno{GotTmn}.

Because $T_{mn}$ is conserved and traceless (the latter due to conformal invariance), one expects that $K^m Q^{\rm tot}_{mn}=0$ where
 \eqn{KmDef}{
  K^m = \begin{pmatrix} v K_1 & K_1 & K_\perp & 0 \end{pmatrix} \,,
 }
and also
 \eqn{TracelessQ}{
  \tr Q^{\rm tot} \equiv -Q^{\rm tot} + Q^{\rm tot}_{11} +
    Q^{\rm tot}_{22} + Q^{\rm tot}_{33} = 0 \,.
 }
The tracelessness condition \eno{TracelessQ} is indeed satisfied, but conservation fails: instead,
 \eqn{Nonconservation}{
  \begin{pmatrix} K^m Q^{\rm tot}_{mn} \end{pmatrix}
    = {iv \over 2} \begin{pmatrix} v & -1 & 0 & 0 \end{pmatrix} \,.
 }
The result \eno{Nonconservation} is independent of $Q_A$, $Q_D$, and $Q_E$: that is, only the last term in the decomposition \eno{QtoQ} fails to be conserved.

The non-conservation \eno{Nonconservation} could have been anticipated.  It is the manifestation of the energy-momentum imparted by the quark to the thermal medium.  The quark is prescribed to travel with constant velocity, so it does not slow down as it loses energy-momentum.  The non-conservation \eno{Nonconservation} should precisely reflect the external force required to keep the quark's momentum from changing, which is minus the drag force \eno{DragForce}.  This argument is formal because the quark's mass is infinite, hence so is its momentum.  But changes in the momentum, and therefore forces, can be finite.  To verify that \eno{DragForce} can be recovered from \eno{Nonconservation}, consider some finite region $V$ of ${\bf R}^3$.  The external force on this region is
 \eqn{Fregion}{
  F^j = {d \over dt} \int_V d^3 x \, \langle T^{0j} \rangle +
    \oint_{\partial V} d^2 a \, n_i \langle T^{ij} \rangle 
    = \int_V d^3 x \, \partial_m \langle T^{mj} \rangle \,.
 }
Here and in the following, $i$ and $j$ are three-dimensional spatial indices, while $m$ and $n$ are $3+1$-dimensional Lorentz indices.  The first term in the middle expression of \eno{Fregion} is the rate of change of momentum in this region, and the second term is the rate of escape of momentum through its boundaries.  Using \eno{FullTmn}, one obtains
 \eqn{Fnext}{
  F^j &= \int_V d^3 x \int {d^3 K \over (2\pi)^3}
   {i \over z_H} K_m \langle T_K^{mj} \rangle
     e^{i \left[ K_1 (x^1-vt) + K_2 x^2 + K_3 x^3 \right]/z_H}  \cr
  K_m &= \begin{pmatrix} -vK_1 & K_1 & K_2 & K_3 \end{pmatrix} \,.
 }
The expression for $K^m$ in \eno{Fnext} is equivalent to \eno{KmDef} except that we have not specialized to $K_2=K_\perp>0$ and $K_3=0$.  Now take the limit where $V$ covers all of ${\bf R}^3$ so as to obtain the force on the whole system.  Performing the $x$ integral first, one obtains
 \eqn{Fobtain}{
  F^j = iz_H^2 \int d^3 K \, K_m \langle T_K^{mj} \rangle
     e^{-i v K_1 t} \delta(K_1) \delta(K_2) \delta(K_3)
    = iz_H^2 \lim_{\vec{K} \to 0} K_m \langle T_K^{mj} \rangle
 }
where we have anticipated that $K_m \langle T_K^{mj} \rangle$ has a smooth limit as $\vec{K} \to 0$.  Indeed, using \eno{GotTmn} and then \eno{Nonconservation} we arrive at
 \eqn{Fone}{
  F^1 = i {\pi T^2 \sqrt{g_{YM}^2 N} \over \sqrt{1-v^2}}
    \lim_{\vec{K} \to 0} K_m Q^{m1} = 
    {\pi \sqrt{g_{YM}^2 N} \over 2} T^2 {v \over \sqrt{1-v^2}}
 }
As promised, this is minus the drag force \eno{DragForce}.

To restore conservation of $T_{mn}$, one may add to it a ``counterterm:''
 \eqn{AddCounterTerm}{
  T_{mn} \to T_{mn} + {\cal T}_{mn}
 }
where, after passing to a co-moving Fourier description,
 \eqn{CounterTermProperty}{
  K^m {\cal T}^K_{mn} = -K^m \langle T^K_{mn} \rangle = 
    -{iv \over 2} {\pi^3 T^4 \sqrt{g_{YM}^2 N} \over
      \sqrt{1-v^2}} 
     \begin{pmatrix} v & -1 & 0 & 0 \end{pmatrix} \,.
 }
A solution to \eno{CounterTermProperty} which is also traceless is
 \eqn{SolvedCounterTerm}{
  \begin{pmatrix} {\cal T}^K_{00} & {\cal T}^K_{01} \\
    {\cal T}^K_{10} & {\cal T}^K_{11} \end{pmatrix} = 
   {iv \over 2} {\pi^3 T^4 \sqrt{g_{YM}^2 N} \over
      (1-v^2)^{3/2}}
    \begin{pmatrix} 1 + v^2 & -2v \\ -2v & 1 + v^2 \end{pmatrix}
 }
with other components of ${\cal T}^K_{mn}$ vanishing.  Using
 \begin{equation}
   \int {d^3 K \over (2 \pi)^3} {1 \over K_1} e^{i \left[ K_1 (x^1 - v t) + K_2 x^2 + K_3 x^3
   \right]/z_H} = i z_H^2 \theta(x^1-vt) \delta(x^2) \delta(x^3) \,,
 \end{equation}
one finds
 \eqn{PositionSpaceCT}{
  \begin{pmatrix} {\cal T}_{00} & {\cal T}_{01} \\
    {\cal T}_{10} & {\cal T}_{11} \end{pmatrix} = 
    -{v \over 2} {\pi T^2 \sqrt{g_{YM}^2 N} \over (1-v^2)^{3/2}}
      \theta(x^1-vt) \delta(x^2) \delta(x^3)
    \begin{pmatrix} 1 + v^2 & -2v \\ -2v & 1 + v^2 \end{pmatrix} \,.
 }
It would be cleaner if ${\cal T}_{mn}$ had delta function support at the location of the quark, but this does not appear to be possible: ${\cal T}^K_{mn}$ would then be analytic in $K_1$ and $K_\perp$, and there are no analytic solutions to \eno{CounterTermProperty}.  The form \eno{PositionSpaceCT} of the counterterm indicates an unphysical ``string,'' wholly in the boundary theory, that pulls forward on the quark to counteract the drag force.

The upshot of this somewhat extended discussion is that the original non-conserved form \eno{QtoQ} captures the dynamics of dissipation and is non-conserved because it leaves out the external motive force that keeps the momentum of the quark from decreasing.

\subsection{Near the horizon}
\label{HORIZON}

Near the horizon of $AdS_5$-Schwarzschild, and for $vK_1 \neq 0$, the leading approximations to solutions to the equations of motion \eno{AAeom} and \eno{DDeom} are
 \eqn{AAhor}{
  A = {e^{-{i v K_1 \over 8} (\pi-\log 4)} \over
    4 \left( 1 - {i v K_1 \over 2} \right)}
    (1-y)^{1-ivK_1/4} +
   U_A (1-y)^{-ivK_1/4} + V_A (1-y)^{ivK_1/4}
 }
 \eqn{DDhor}{
  \begin{pmatrix} D_1 \\ D_2 \end{pmatrix} &=
   v^2 \cos^2 \theta {e^{-{i v K_1 \over 8} (\pi - \log 4)} \over
    4 \left( 1 - {i v K_1 \over 4} \right)^2}
    \begin{pmatrix} 1-y \\ s_{D_2} \end{pmatrix}
    (1-y)^{-ivK_1/4} +
    T_D^{(1)} \begin{pmatrix} v^2 \cos^2 \theta \\ 1 \end{pmatrix}
   \cr &\qquad{} +
    T_D^{(2)} \begin{pmatrix} 1 \\ t_{D_2} \end{pmatrix} (1-y) +
    U_D \begin{pmatrix} u_{D_1} (1-y) \\ 1 \end{pmatrix}
      (1-y)^{-ivK_1/4}  \cr &\qquad{} +
    V_D \begin{pmatrix} v_{D_1} (1-y) \\ 1 \end{pmatrix}
      (1-y)^{ivK_1/4}
 }
 \eqn{DDhorExtrasDefs}{
  s_{D_2} &= {i \over v K_1} \left( 1 - {ivK_1 \over 4}
    \right) +
   {4 K^2 \over (vK_1)^4} \left( 1 - {ivK_1 \over 4} \right)^2  \cr
  t_{D_2} &= {K^2 \over 16 + (vK_1)^2} \qquad
  u_{D_1} = -{ivK_1 \over 1 - {ivK_1 \over 4}} \qquad
  v_{D_1} = {ivK_1 \over 1 + {ivK_1 \over 4}}
 }
 \eqn{EEhor}{
  \begin{pmatrix} E_1 \\ E_2 \\ E_3 \\ E_4 \end{pmatrix} &=
   {i v e^{-{i v K_1 \over 8} (\pi - \log 4)} \over
    2 K_1 \left( 1 + {i v K_1 \over 2} \right) }\begin{pmatrix} 1 \\ 0 \\ 1 \\ 0 \end{pmatrix} +
   T_E^{(1)} \begin{pmatrix} 1 \\ -\frac{1}{2} \\ 1 \\ -2
   \end{pmatrix} +  T_E^{(2)}\begin{pmatrix} \frac{3}{2} K_1^2 v^2
   \\ {1 \over 4} K^2 (1-y^4) \\ 0 \\ 0 \end{pmatrix} + \frac{T_E^{(3)}}{\sqrt{1-y}}
   \begin{pmatrix} 3 (4 + K_1^2 v^2) \\ 0 \\ 4 K^2 (1-y) \\ -8 K^2
   (1-y) \end{pmatrix}\cr &{} + T_E^{(4)} \begin{pmatrix} K_1^2 v^2
   \log(1-y) \\ 2 \\ 0 \\ 8 \end{pmatrix} + T_E^{(5)} \begin{pmatrix}
   1 \\ 0 \\ 0 \\ \frac{8}{3} K^2 (1-y) \end{pmatrix}  +
   \frac{T_E^{(6)}}{\sqrt{1-y}} \begin{pmatrix} 3(4 + K_1^2 v^2) \\
   0 \\ 4(6 + K^2)(1-y) \\ -8 K^2 (1-y) \end{pmatrix} \cr &{}+ U_E
   (1-y)^{-i v K_1/4} \begin{pmatrix} 0 \\ 0 \\ 0 \\ 1 \end{pmatrix}+ V_E
   (1-y)^{i v K_1/4} \begin{pmatrix} 0 \\ 0 \\ 0 \\ 1 \end{pmatrix}
    \,.
 }
In these solutions, $U_X$, $V_X$, and $T_X^{(i)}$ are integration constants.  Near-horizon solutions to the $B$ and $C$ equations are identical, respectively, to the $D$ and $A$ solutions \eno{DDhor} and \eno{AAhor}, except that the particular solutions are zero.

For each set of equations, the solution multiplied by $U_X$ is infalling (meaning that gravitons are falling into the black hole), while the solution multiplied by $V_X$ is outfalling.  The solutions multiplied by $T^{(i)}_X$ are neither infalling nor outfalling but can  be categorized by their regularity properties at the horizon.  The standard boundary condition imposed at a black hole horizon is that outfalling modes must vanish: $V_X=0$ for $X=A$, $B$, $C$, $D$, and $E$.

The constraint equations \eno{DDconstraint} and \eno{EEconstraint} imply
 \eqn{ConstraintsImply}{
  T_D^{(2)} = T_E^{(3)} = T_E^{(4)} = T_E^{(5)} = 0 \,.
 }
The solutions in \eno{DDhor} and \eno{EEhor} multiplied by $T_D^{(1)}$, $T_E^{(1)}$, and $T_E^{(2)}$ are in fact exact solutions to the equations of motion for all $y$.  Note that the exact solutions do not overlap with the ones removed by \eno{ConstraintsImply}.  This suggests that in the coupled systems of equations of motion and constraints for $D_i$ and $E_i$, it may be possible to make further reductions of order.  We have not pursued this avenue, but it might facilitate future numerical studies.

To understand the boundary value problem that determines $\langle T^K_{mn} \rangle$, it is useful first to review the counting of integration constants, constraints, and boundary conditions:
 \begin{itemize}
  \item The ten second order equations of motion have $20$ constants of integration which must be fixed in order to specify a unique solution.
  \item Ten constants of integration are fixed by requiring $R_X=0$ at the boundary (no deformations of the gauge theory).  The other ten are the $Q^{\rm tot}_X$, which are linear combinations of entries of $\langle T^K_{mn} \rangle$.
  \item Five relations among the $Q^{\rm tot}_X$ follow from imposing the constraints at the boundary.
  \item Five more boundary conditions must be imposed at the horizon to suppress the outfalling solutions.
 \end{itemize}
Evidently, the number of constraints plus boundary conditions equals the number of integration constants in the equations of motion.  So the boundary value problem is well posed.  All the integration constants are complex, and the constraints and boundary conditions are too.

Similar counting of integration constants can be done after dividing the equations into the decoupled sets, $A$ through $E$.  Let us include this counting in a summary of the numerical algorithm we used.  In the $A$ set, one must impose $R_A=0$.  If we supply in addition an {\it ad hoc} value $q$ for $Q^{\rm tot}_A$, then Cauchy data has been specified at the boundary.  More precisely, approximate Cauchy data can be specified at a finite but small value $y=y_0$ by setting $R_A=0$ and $Q^{\rm tot}_A=q$ equal to its {\it ad hoc} value and using \eno{XsetBoundary} to determine $A(y_0)$ and $A'(y_0)$.  The second order equation of motion \eno{AAeom} can then be integrated numerically to $y=y_1$, where $y_1$ is close to $1$.  The numerical solution can then be fit to the asymptotic form \eno{AAhor}, and values of $U_A$ and $V_A$ can be extracted.  Because all the equations are affine (meaning linear with inhomogeneous terms), $V_A$ (as well as $U_A$) is an affine function $V_A(q)$ of the {\it ad hoc} value we supplied for $Q^{\rm tot}_A$.  The equation $V_A(q)=0$ may easily be solved for the physically meaningful value of $Q^{\rm tot}_A$.

For the $D$ and $E$ sets, the situation is only slightly more complicated.  After setting $R_X=0$ (see \eno{XsetBoundary}) and imposing the constraint (for $D$) or constraints (for $E$), there is only one degree of freedom left at the boundary, which we can fix by supplying an ad hoc value for the quantity $Q^{\rm tot}_D$ or $Q^{\rm tot}_E$ that enters \eno{QtoQ}.  Cauchy data for the equations of motion can be generated at $y=y_0$, and after numerically solving the equations of motion, the integration constants $T^{(i)}_X$, $V_X$, and $Q^{\rm tot}_X$ can be extracted by matching numerics to horizon asymptotics at $y=y_1$.  To determine $Q^{\rm tot}_D$ or $Q^{\rm tot}_E$, one solves an affine equation $V_X(q)=0$.  It would have been numerically more efficient to eschew one of the equations of motion in the $D$ set and all but one in the $E$ set in favor of the first order constraint equations.  But we found it a useful check of numerical accuracy to evaluate at $y=y_1$ the $T^{(i)}_X$ which are required by \eno{ConstraintsImply} to vanish.

The method of obtaining an affine function at the horizon by first specifying Cauchy data at the boundary was described in \cite{Teaney:2006nc} for graviton perturbations in $AdS_5$-Schwarzschild in the absence of the trailing string.

\subsection{Large $K$ behavior}
\label{NEARFIELD}

The large $K$ behavior of $Q^{\rm tot}_{mn}$ is dominated by the near-field of the quark.  For $v=0$, this field is entirely color-electric, and it is perfectly Coulombic because of the conformal symmetry of ${\cal N}=4$ super-Yang-Mills theory.  These observations alone, together with the radial symmetry and the conservation and tracelessness conditions, fix the $v=0$ form of $Q^{\rm tot}_{mn}$ up to an overall prefactor:
 \eqn{GotNearField}{
  Q^{\rm near}_{mn} =
   -{\pi \over 24 |\vec{K}|}
   \begin{pmatrix} 2 \vec{K}^2 & 0 & 0 & 0 \\
     0 & K_2^2 + K_3^2 & -K_1 K_2 & -K_1 K_3  \\
     0 & -K_1 K_2 & K_1^2 + K_3^2 & -K_2 K_3  \\
     0 & -K_1 K_3 & -K_2 K_3 & K_1^2 + K_2^2 \end{pmatrix} \,.
 }
In section~\ref{NORMALIZE} we explain how to fix the overall prefactor in \eno{GotNearField}.

$Q^{\rm near}_{mn}$ for $v \neq 0$ can be obtained by applying a Lorentz boost to \eno{GotNearField}.  After this is done, one may define
 \eqn{QKdef}{
  Q^K_{mn} = Q^{\rm tot}_{mn} - Q^{\rm near}_{mn} \,.
 }
Then $Q^K_{mn} \to 0$ as $K \to \infty$.\footnote{Actually, $Q^K_{mn}$ can be arranged to have an arbitrary $\vec{K}$-independent limit for large $K$ by adjusting the choice of particular solution.  This corresponds to adjusting a subtraction scheme for the infinite self-energy of the external quark.  The form of solutions specified in \eno{BdySolns}, where the particular solution is assumed not to have a quartic term, is a sort of holographic minimal subtraction scheme.}  More useful in section~\ref{NUMERICS} will be the equivalent forms
 \eqn{QXdef}{
  Q^K_X = Q^{\rm tot}_X - Q^{\rm near}_X \qquad
    X = A,D,E
 }
where
 \eqn{QXnear}{
  Q^{\rm near}_A = Q^{\rm near}_D =
    {\pi K \over 16} \sqrt{1-v^2\cos^2\theta}  \qquad
  Q^{\rm near}_E = {\pi K \over 24}
    {2 + v^2 (1-3\cos^2\theta) \over
      \sqrt{1-v^2\cos^2\theta}} \,.
 }
Recall that $Q_D = Q_{D_1}$ and $Q_E = Q_{E_1}$.  To derive \eno{QXnear} one must compare the conserved terms in \eno{QtoQ} with the Lorentz-boosted version of \eno{GotNearField}, with $K_3=0$ and $K_2=K_\perp$ in the rest frame of the thermal plasma.

\subsection{Normalizing the near field}
\label{NORMALIZE}

The prefactor can be fixed by observing that the equations \eno{LinODEs} for $v=0$, $h=1$, and $\xi=0$, as appropriate for a static string in $AdS_5$, can be solved for $K_\perp=0$ by setting all the $H_{mn} = 0$ except for $H_{00}$, $H_{11} = 2y^3/9$, and $H_{22} = H_{33} = H_{00}/2 = -2f/3$, where $f(y)$ satisfies
 \eqn{hzzBest}{
  \left[ \partial_y^2 - {3 \over y} \partial_y - K_1^2 \right]
    f = y \,.
 }
This is precisely the equation satisfied by the $K_\perp = 0$ Fourier modes $\tilde\phi_K$ of the dilaton sourced by the same static string configuration: see~(17) of~\cite{Friess:2006aw}.  Although \eno{hzzBest} is non-trivial to solve directly, position space methods are available to extract the dilaton profile \cite{Danielsson:1998wt}.  From them one can Fourier transform back to find $B_K = \pi |K_1|/16$, where $B_K$ is defined through the asymptotic behavior
 \eqn{dilAsympt}{
  \tilde\phi_K = -{y^3 \over 3} + B_K y^4 \,.
 }
From $H_{00} = -4f/3$ it follows that $Q^{\rm near}_{00} = -4B_K/3$; hence the prefactor in \eno{GotNearField}.  Comparing (23) of \cite{Friess:2006aw} to \eno{GotTmn} of the current paper, one may conclude that
 \eqn{TvsO}{
  \langle T_{00} \rangle = {4 \over 3} \langle {\cal O}_{F^2}
    \rangle \,.
 }
This is a positive quantity because ${\cal O}_{F^2} \sim \tr \vec{E}^2$ for the static quark.  We do not know how to account for the factor of $4/3$ in \eno{TvsO}.

It is instructive to examine the same static quark solution using the equations \eno{AAdef}-\eno{EEconstraint}.  Taking $h=1$ and $\xi=0$ in these equations poses no difficulties.  $K_\perp = 0$ means $\theta=0$, which appears to lead to difficulties in~\eno{AAdef} and~\eno{DDdef} (the definitions of $A$, $D_1$, and $D_2$ in terms of $H_{mn}$).  But the inverse relations expressing $H_{mn}$ in terms of the $ABCDE$ variables are entirely non-singular in the limit $K_\perp \to 0$: they read
 \eqn{HfromABCDE}{\seqalign{\span\TL & \span\TR &\qquad \span\TL & \span\TR}{
  H_{00} &= -{2 \over 3} E_1 + {2 \over 3} E_3 &
   H_{01} &= 2v E_2  \cr
  H_{02} &= 0 & H_{11} &= {2 \over 3} E_3 - {2 \over 3} E_4  \cr
  H_{12} &= 0 & H_{22} &= {2 \over 3} E_3 + {1 \over 3} E_4  \cr
  H_{33} &= {2 \over 3} E_3 + {1 \over 3} E_4 \,.
 }}
We have omitted expressions for the ${\bf Z}_2$-odd components of the metric in terms of $B_i$ and $C$ because all these quantities vanish in the solution we're interested in.  The equations of motion and constraints for $A$, $D_i$, and $E_i$ are also non-singular in the limit $K_\perp \to 0$.  The equations of motion and constraints for $A$ and $E_i$ are non-singular if one additionally takes $v \to 0$, but the constraint for $D_i$ is not.  Therefore in \eno{AstaticEOMS}-\eno{EstaticEOMS} we partially quote and partially solve the equations of motion and constraints for $A$ and $E_i$ after having set $v=0$, but for $D_i$ we keep $v$ finite.
 \eqn{AstaticEOMS}{
  \left[ \partial_y^2 - {3 \over y} \partial_y - K_1^2
    \right] A = y
 }
takes precisely the same form as \eno{hzzBest}.
 \eqn{DstaticConstraints}{
  D_1 = d_1 + {i y^4 \over 4 K_1} {v \over 1-v^2} \qquad
  D_2 = d_1 + {i y^4 \over 4 K_1} {1 \over v(1-v^2)}
 }
is the general solution of the $D$ constraint consistent with the requirement that $D_i \to 0$ as $y \to 0$.  The function $d_1$ satisfies
 \eqn{DstaticEOMS}{
  \left[ \partial_y^2 - {3 \over y} \partial_y +
   K_1^2 (1-v^2) \right] d_1 = y \,,
 }
which takes the same form as \eno{hzzBest} except for the replacement $K_1 \to K_1 \sqrt{1-v^2}$.
 \eqn{EstaticConstraints}{
  E_1 = {4 \over 3} e_1 + {y^3 \over 9} \qquad
  E_2 = -E_3 = {2 \over 3} e_1 - {y^3 \over 9} \qquad
  E_4 = -{2 \over 3} e_1 - {2y^3 \over 9}
 }
is the general solution of the $E$ constraint consistent with the requirement that $E_i \to 0$ as $y \to 0$.  The function $e_1$ satisfies
 \eqn{EstaticEOMS}{
  \left[ \partial_y^2 - {3 \over y} \partial_y - K_1^2
    \right] e_1 = y \,,
 }
which again is precisely the same form as \eno{hzzBest},

Because \eno{HfromABCDE} involves only the $E_i$, it was superfluous to explicitly solve the $A$ and $D_i$ equations in \eno{AstaticEOMS}.  But it is a worthwhile check to ensure that the quantities $D_i$, though singular in the limit $v \to 0$ (as well as in the limit $K_1 \to 0$), cause no problems for $H_{mn}$.  Indeed one recovers the results for $H_{mn}$ stated briefly around \eno{hzzBest}.

\subsection{Small $K$ behavior}
\label{SMALLK}

A series expansion in small $K$ allows some progress to be made on
solving the equations of motion.  We will first focus on the
simplest case, namely the $A$ equation. Plugging
 \eqn{Aexpand}{
  A = \alpha_0 + K \alpha_1 + K^2 \alpha_2 + \ldots
 }
into \eno{AAeom}, one can find the differential equations satisfied
by each $\alpha_j$.  The first three are:
 \eqn{aIterative}{
  {y^3 \over h} \partial_y {h \over y^3} \partial_y \alpha_0 &=
    {y \over h}  \cr
  {y^3 \over h} \partial_y {h \over y^3} \partial_y \alpha_1 &=
    -i {y \over h} {\cos\theta \xi \over z_H}  \cr
  {y^3 \over h} \partial_y {h \over y^3} \partial_y \alpha_2 &=
    -{y \over h} \left( {\cos\theta \over z_H} \right)^2 -
    {v^2 \cos^2 \theta - h \over h^2} \alpha_0  \,.
 }
Evidently, these equations are solvable through repeated
integration.  Integrating $\alpha_0$ is easy, and after matching to
the boundary asymptotics with $R_A=0$ and the horizon asymptotics
with $V_A=0$, (both suitably expanded in $K$) one obtains
 \eqn{GotAlphaZero}{
  \alpha_0 = -{1 \over 2} \log(1+y) - {1 \over 4} \log(1+y^2) +
    {i \over 4} \log {1-iy \over 1+iy} = -{y^3 \over 3} +
      {y^4 \over 4} + O(y^7) \,,
 }
which implies $Q^{\rm tot}_A = 1/4 + O(K)$.  Higher order
corrections to $ Q^{\rm tot}_A$ can be found by solving the
corresponding differential equations for each of the $\alpha_j$'s,
and matching the solutions to the horizon and boundary asymptotics.
To order $O(K^2)$ we have obtained
 \eqn{QAseries}{
  Q^{\rm tot}_A &= {1 \over 4} - {i \log 2 \over 8} vK_1 +
   {K^2 \over 192} (6\pi - 12 \log 2) \Big[ \sin^2\theta
     \cr
   &\qquad\qquad {} +
    \left( 1-v^2 + v^2 {\pi^2 - 12 (\log 2)^2 \over
      6\pi - 12 \log 2} \right) \cos^2 \theta \Big] +
      O(K^3) \,.
 }
A similar analysis can be carried out for the $D$ and $E$ sets.
For the $D$ set, for example, one first writes down series expansions
in $K$ for the second order differential equations \eno{DDeom}, and
solves for the corrections to $D_1$ and $D_2$ iteratively.  By
imposing the constraint \eno{DDconstraint} and by matching, at each
order, the small $K$ solution to the boundary asymptotics with
$R_{D_1}=R_{D_2}=0$ and to the horizon asymptotics with $V_D=0$, one
can then solve for the four integration constants and the
corresponding corrections to $Q^{\rm tot}_{D_1}$, $Q^{\rm tot}_{D_2}$, $U_D$, $T^{(1)}_D$, and $T^{(2)}_D$.  To linear order in $K$, we find
 \eqn{QDseries}{
  Q^{\rm tot}_D &= -{i \sec\theta \over 4vK} -
    {\sec^2\theta - 4v^2 \over 16 v^2} +
    {iK \sec^3\theta \over 64 v^3} \big[ 1 +
     (2 \log 2) v^2 \cos^2 \theta  \cr &\qquad\qquad {} - (8 \log 2)
      v^4 \cos^4 \theta \big] +
   O(K^2) \,.
 }
Similarly, for the $E$ set, we find
 \eqn{QEseries}{
  Q^{\rm tot}_E &= {3 i v (1+v^2) \cos\theta \over 2 K
     \left(1 - 3 v^2 \cos^2 \theta \right)} - 
    {3 v^2 \cos^2 \theta \left[ 2 + v^2 
      \left( 1 - 3 \cos^2\theta \right) \right]
      \over 2 \left( 1 - 3 v^2 \cos^2\theta \right)^2} +
    O(K)  \cr
   &= {3iv (1+v^2) \cos\theta \over 2K}
    {1 \over (1-3v^2\cos^2 \theta)\left( 1 - {ivK \cos\theta \over
      1+v^2} \right) - ivK \cos\theta} + O(K) \,.
 }
The striking feature of the expression is the singular behavior at $\theta = \cos^{-1}(1/v\sqrt{3})$, which is the Mach angle.  From this we may conclude that there is a sonic boom in the thermal medium involving large amplitude but small momentum fields.  In the second expression, we have shown how the $O(1)$ term may be ``resummed'' into the leading $O(1/K)$ expression so as to blunt the singularity into a form resembling a Lorentzian lineshape.

\section{Results of numerics}
\label{NUMERICS}

 \begin{figure}
  \centerline{\includegraphics[width=5in]{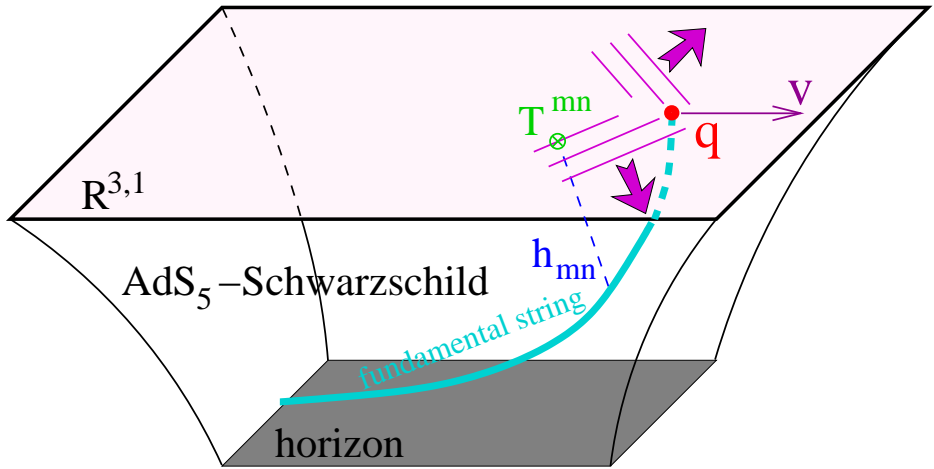}}
  \caption{The $AdS_5$-Schwarzschild background is part of the near-extremal D3-brane, which encodes a thermal state of ${\cal N}=4$ supersymmetric gauge theory \cite{Gubser:1996de}.  The external quark trails a string into the five-dimensional bulk, representing color fields sourced by the quark's fundamental charge and interacting with the thermal medium.}\label{fig:wake}
 \end{figure}
 
Let us briefly recap the five-dimensional gravitational calculation that has been our main focus so far.  The trailing string of \cite{Herzog:2006gh,Gubser:2006bz} sources the graviton, which propagates classically in $AdS_5$-Schwarzschild with purely infalling boundary conditions at the black hole horizon.  The graviton's behavior near the boundary of $AdS_5$-Schwarzschild determines $\langle T_{mn} \rangle$ in the boundary gauge theory.  Thus $\langle T_{mn} \rangle$ is a shadow  (other authors might prefer the term ``hologram'') of the trailing string.  See figure~\ref{fig:wake}.
 \begin{figure}
  \centerline{\includegraphics[width=5.5in,angle=270]{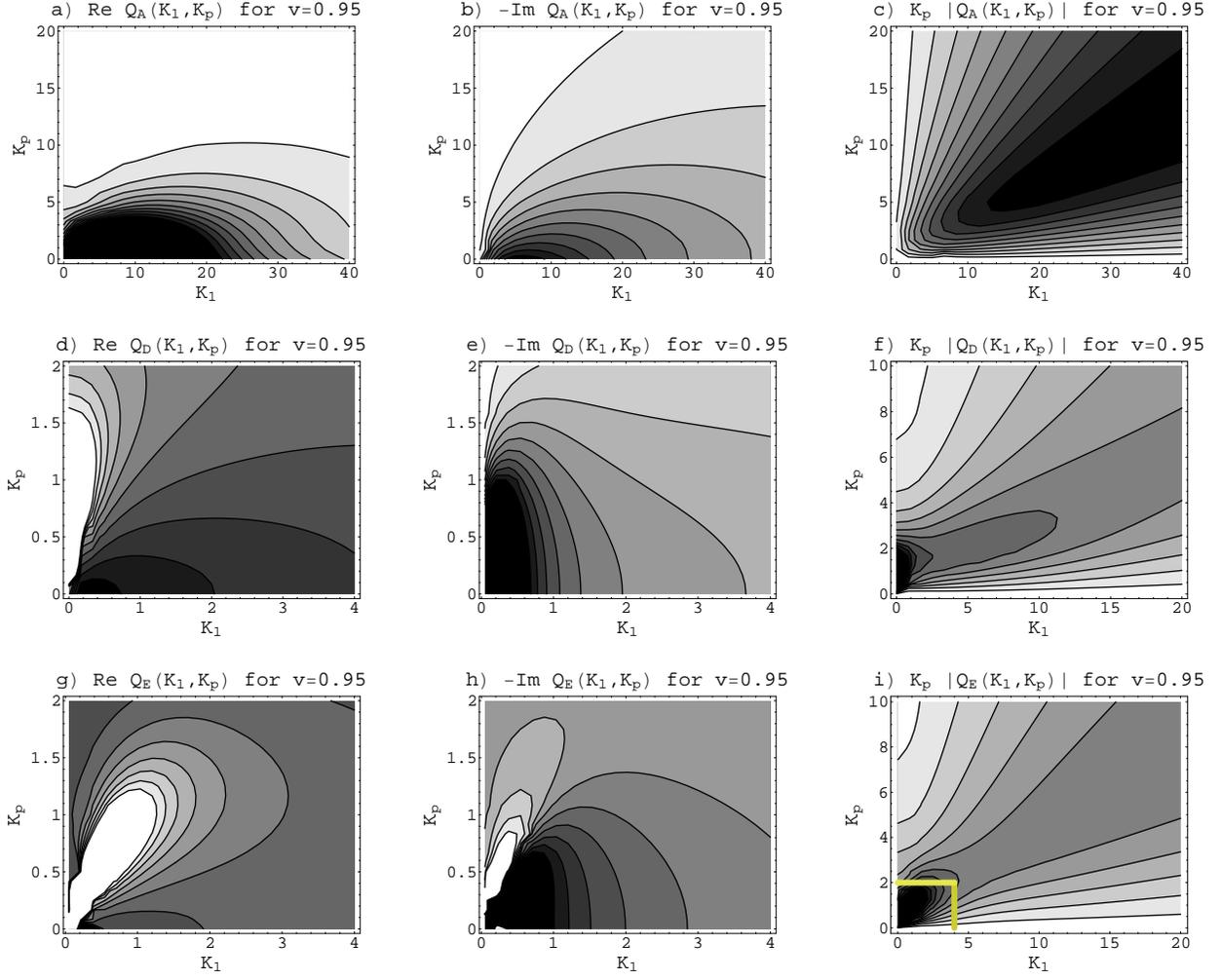}}
  \caption{Contour plots of $Q^K_A$, $Q^K_D$, and $Q^K_E$ for $v=0.95$.  The darker regions are more positive.  All components of $\langle T^K_{mn} \rangle$ can be deduced from $Q^K_A$, $Q^K_D$, and $Q^K_E$ using \eno{QtoQ}, \eno{GotTmn}, and \eno{QXdef}.  All three $Q^K_X$ go to zero at large $K$.  The momentum vector $\vec{K} = \vec{k}/\pi T$ can be read in ${\rm GeV}/c$ if one chooses $T = 318\,{\rm MeV}$: see \eno{SetT}.  The range of momenta in each plot was chosen to show the most distinctive structures.  The region outlined in gold in (i) is plotted in more detail in figure~\ref{fig:LobeFig}c.}\label{fig:ADEfig}
 \end{figure}

Our aim is to describe $\langle T_{mn} \rangle$ in the boundary theory.  We will focus on Fourier coefficients $Q^K_X$ for $X=A$, $D$, and $E$.  As reviewed at the end of section~\ref{INTRODUCTION}, these quantities are Fourier coefficients of linear combinations of entries of $\langle T_{mn} \rangle$ with a near-field subtraction.  Our numerical algorithm is outlined at the end of section~\ref{HORIZON}.  It was implemented primarily using Mathematica's {\tt NDSolve}.  To achieve good accuracy, it was necessary to develop asymptotic power series solutions to a considerably higher order than shown in sections~\ref{BOUNDARY} and~\ref{HORIZON}.  Experience as well as common sense suggest that large $K$ regions become more numerically challenging.  We believe we have adequately met this challenge, partly by allowing arbitrarily many steps in {\tt NDSolve} and calculating with a working precision of $30$ digits (i.e.~roughly twice the standard double precision of modern PC's).  Another challenging region is small $K_1$, where the outfalling and infalling solutions are nearly constant until $y$ is very close to $1$.  Experience suggests that at most a narrow region with $K_1 \ll K_\perp$ is problematic. \begin{figure}
  \centerline{\includegraphics[width=4in,angle=270]{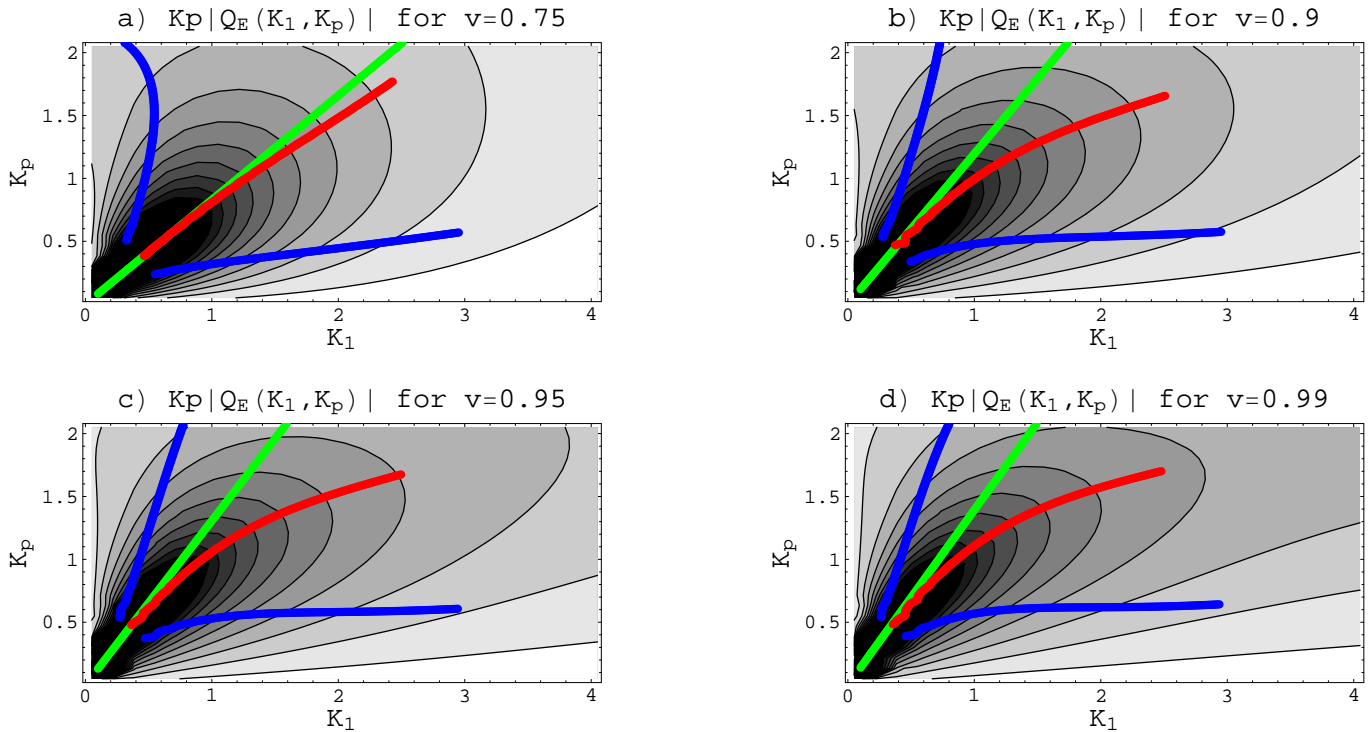}}
  \caption{Contour plots of $K_\perp |Q^K_E|$ for various values of $v$.  $Q^K_E$ is proportional to the $K$-th Fourier component of the energy density after a near-field subtraction: see~\eno{QtoQ}, \eno{GotTmn}, and \eno{QXdef}.  The phase space factor $K_\perp$ arises in Fourier transforming back to position space.  The green line shows the Mach angle.  The red curve shows where $K_\perp |Q^K_E|$ is maximized for fixed $K = \sqrt{K_1^2+K_\perp^2}$.  The blue curves show where $K_\perp |Q^K_E|$ takes on half its maximum value for fixed $K$.}\label{fig:LobeFig}
 \end{figure}
 \begin{figure}
  \centerline{\includegraphics[width=5in,angle=270]{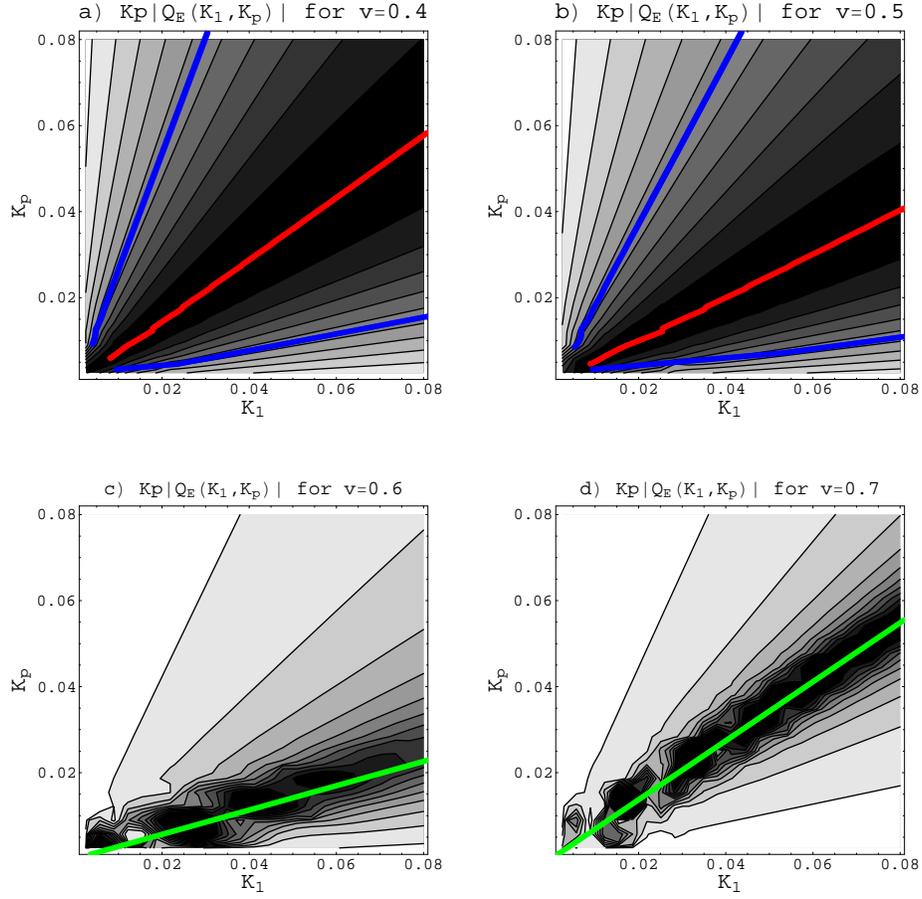}}
  \caption{Contour plots of $K_\perp |Q^K_E|$ for various values of $v$ at low momenta.  The green line shows the Mach angle.  The red curve shows where $K_\perp |Q^K_E|$ is maximized for fixed $K = \sqrt{K_1^2+K_\perp^2}$.  The blue curves show where $K_\perp |Q^K_E|$ takes on half its maximum value for fixed $K$.}\label{fig:BoomFig}
 \end{figure}

We found excellent agreement between the numerically computed $Q^{\rm tot}_X$ and the analytical approximations $Q^{\rm near}_X$ for large $K$ and for $X=A$, $D$, and $E$.  Only the subtracted quantities $Q^K_X=Q^{\rm tot}_X-Q^{\rm near}_X$ appear in the plots in figures~\ref{fig:ADEfig}-\ref{fig:LobeFig}.   For $Q^K_A$ and $Q^K_D$, we also found excellent agreement with the small $K$ asymptotics \eno{QAseries} and \eno{QDseries} in the expected ranges.  For $Q^K_E$, the nearly singular behavior near the origin is difficult for numerics to capture.  This difficulty shows up in the ragged contours in figure~\ref{fig:BoomFig}c,d.  The problem is not numerical error in evaluations of $Q^K_E$ at individual points; rather, the ragged contours in figure~\ref{fig:BoomFig}c,d are due mostly to imperfect interpolations over a grid of limited resolution.  Indeed, individual evaluations of $Q^K_E$ for $K = \sqrt{K_1^2 + K_\perp^2} = 0.08$ agree with the small $K$ asymptotics \eno{QEseries} at the level of about a percent.  A high-resolution plot of $K_\perp |Q^K_E|$ at $K=0.08$ is shown in figure~\ref{fig:QEfixedK}a.  In this plot, the results of numerics are visually indistinguishable from the analytic form \eno{QEseries}.  Even in the coarser-grained numerical evaluations of $Q^K_E$ shown in figure~\ref{fig:BoomFig}, agreement with \eno{QEseries} was good a distance $\delta K \sim 0.015$ away from the central ridge.

Two qualitative features visible in figures~\ref{fig:ADEfig}-\ref{fig:LobeFig} are worthy of note.  The first is the high momentum ridges, which are most distinctive in figure~\ref{fig:ADEfig}c.  This is the same feature that was noted in \cite{Friess:2006aw}; indeed, $Q^K_A$ of this paper is identical to $B_K$ of \cite{Friess:2006aw}.  High momentum ridges are also present in $Q^K_D$ and $Q^K_E$.  For $v=0.95$ and $v=0.99$, we find empirically that $Q^K_A \approx 4 Q^K_D$ on the high momentum ridges.  A more approximate relation for a similar region of momenta is $Q^K_E = 3 Q^K_D$.

The second feature worthy of note is the sharp structures at low momentum in figures~\ref{fig:ADEfig}g,h,i.  A more detailed view of these structures is shown in figure~\ref{fig:LobeFig}.  As we will discuss in section~\ref{DISCUSSION}, the lobes in figure~\ref{fig:LobeFig} are suggestive of high angle emission of particles in energy ranges accessible to experiments at RHIC\@.  The lobes become narrower as one passes to small $\vec{K}$, corresponding to momenta much less than the temperature: see figure~\ref{fig:BoomFig}c,d.  Low momentum is the hydrodynamic limit, so it is gratifying to see a highly directional feature corresponding to a sonic boom.  Figures~\ref{fig:BoomFig}c,d thus serve as visual confirmation of the appearance of a sonic boom that we anticipated based on \eno{QEseries}.  Figures~\ref{fig:BoomFig}a,b show what happens when the velocity of the quark falls below the speed of sound $1/\sqrt{3} \approx 0.577$ in the thermal medium.  Evidently, there is still directional emission, but it becomes abruptly less focused.  The peak amplitude also decreases abruptly.  Intriguingly, the drag force \eno{DragForce} behaves completely smoothly as one passes through $v=1/\sqrt{3}$.

Readers wishing to examine our results more quantitatively are referred to \cite{QXwebsite}.

\section{Discussion}
\label{DISCUSSION}

Because all the calculations in this paper were carried out in the framework of five-dimensional supergravity coupled to a classical string, the gauge theory results are accurate only to leading order in large $N$ and large $g_{YM}^2 N$.  Large 't~Hooft coupling is largely inaccessible to standard techniques of finite-temperature quantum field theory, with the important exception of lattice gauge theory.  But finite-temperature lattice methods are not well-adapted to real-time dissipative phenomena, in contrast to AdS/CFT, which provides ready access to both static and dissipative properties.  Moreover, the AdS/CFT prescription for computing gauge theory observables is conceptually the same at all energy scales, giving some advantage over hydrodynamical approximations that are best justified in the infrared limit.  So AdS/CFT occupies a unique niche in the range of tools available for understanding strongly coupled gauge theories at finite temperature.  Its principal drawback, at least as we use it in this paper, is that the dual gauge theory is $SU(N)$ ${\cal N}=4$ super-Yang-Mills, which in some ways is quite distant from real-world QCD\@.  Within the limitations that we have described, the calculation of $\langle T_{mn} \rangle$ provides a fairly comprehensive description of dissipation from the heavy quark.  All possible gauge interactions are included, in particular secondary interactions with the thermal medium of energetic particles radiated from the quark.

For the sake of definiteness, let us set
 \eqn{SetT}{
  T = {1 \over \pi}\,{\rm GeV} = 318\,{\rm MeV} \,.
 }
We understand this number to be in the upper range of temperatures for the quark-gluon plasma (QGP) produced at RHIC\@.  It is a convenient choice for us because the $K_1$ and $K_\perp$ axes in figures~\ref{fig:ADEfig}, \ref{fig:LobeFig}, and \ref{fig:BoomFig} can then be read in units of ${\rm GeV}/c$.

In \cite{Friess:2006aw} we suggested that the high momentum ridges might be evidence that the strongly coupled thermal medium enhances fragmentation near the kinematic limit.  But one of our warnings was that one should compute $\langle T_{mn} \rangle$ before making definitive statements.  In light of the pronounced directional peak in $K_\perp |Q^K_E|$ at low $K$, we are inclined to regard the high momentum ridges as less immediately important to attempts to compare string theory calculations to recent experimental results on jet-quenching.  It is plausible that these ridges are the expression in Fourier space of a sharp ``prow'' of color fields supported near the quark.  Moreover, we must bear in mind that at the typical energy scale $E \gsim 10\,{\rm GeV}$ where the high momentum ridges are pronounced, QCD is no longer particularly strongly coupled, so there is less justification for a connection with the supergravity approximation in AdS/CFT\@.\footnote{When hadron pair correlators are plotted with higher momenta windows for the hadrons, we understand from an experimental colleague that the away side peak reappears \cite{JacakPrivate}.  The high momentum ridges might be relevant to such correlators: forward emission is indeed what they imply.  But it is perhaps more plausible to attribute the reappearance of the away-side peak to away-side partons that have enough energy to punch through the QGP with only modest deflection.} 

Our most striking results are the directional lobes in $K_\perp |Q^K_E|$, as seen in figure~\ref{fig:LobeFig}.  Recall that the factor of $K_\perp$ is appropriate because it is the measure factor arising in a Fourier transform back to position space after the azimuthal integral is performed.  Recall also that $Q^K_E$ is directly proportional to the $K$-th Fourier coefficient $\langle T^K_{00} \rangle$ of the energy density with the Coulombic near field subtracted away, whereas $Q^K_A$ and $Q^K_D$ are combinations of components of $\langle T^K_{mn} \rangle$ and non-conserved terms: see \eno{QtoQ}, \eno{GotTmn}, and \eno{QXdef}.  From figure~\ref{fig:LobeFig} we conclude that in strongly coupled ${\cal N}=4$ gauge theory at finite temperature, directional emission from a hard probe is present, but not sharply focused, at momenta several times the temperature.  This seems to us an intuitively appealing conclusion: Rescattering effects broaden the directionality of the ``wake'' in Fourier space.
 \begin{figure}
  \centerline{\includegraphics[width=6in]{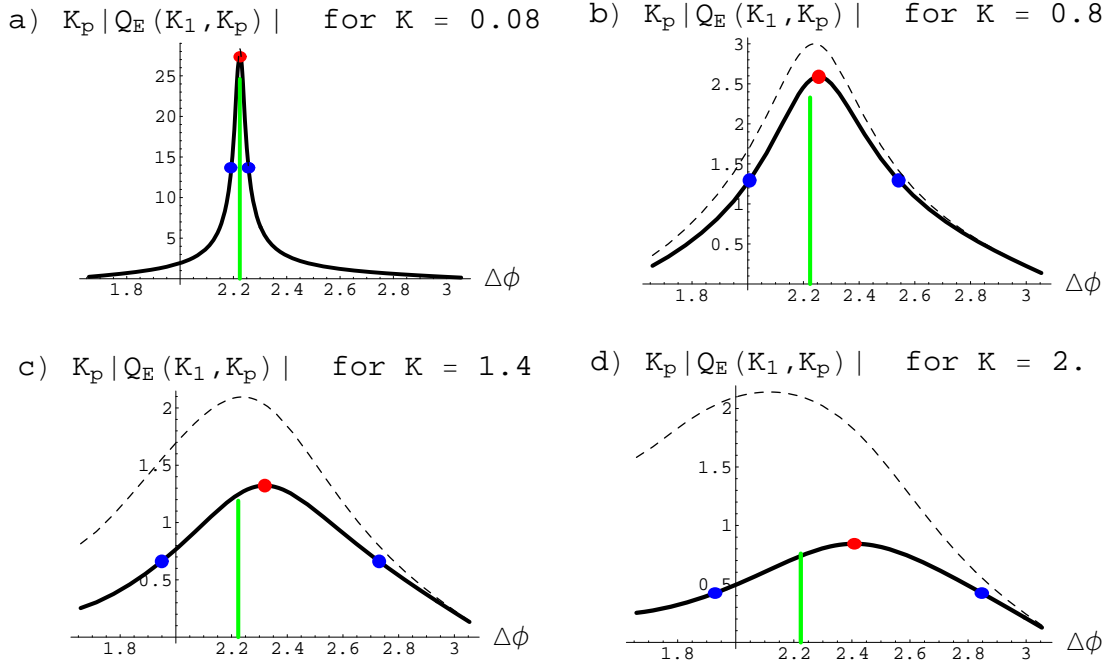}}
  \caption{$K_\perp |Q^K_E|$ at fixed $K = \sqrt{K_1^2+K_\perp^2}$ as a function of angle, for $v=0.95$ and for various values of $K$.  To facilitate comparison with di-jet hadron pair correlations, we have parameterized the angle as $\Delta\phi = \pi-\theta$, where $\theta=\tan^{-1} K_\perp/K_1$.  With the usual assignment $T = 318\,{\rm MeV}$ (see \eno{SetT}), $K$ can be read in units of ${\rm GeV}/c$.  In each plot, the solid curve is from numerics; the dashed curve is the analytical approximation \eno{QEseries}; the green line indicates the Mach angle; the red dot is at the maximum of $K_\perp |Q^K_E|$; and the blue dots indicate the points where $K_\perp |Q^K_E|$ is half of its peak value.}\label{fig:QEfixedK}
 \end{figure}

In figure~\ref{fig:QEfixedK} we show a small sampling of our numerical results in a format more suggestive of a comparison with experimental results \cite{Wang:2004kf,Adler:2005ee} on the splitting of the away side peak in di-jet hadron pair correlations.  What we find attractive is that at momenta comparable to the window $1\,{\rm GeV}/c < p_T < 2.5\,{\rm GeV}/c$ of transverse momenta for the partner hadrons, there are broad peaks in the Fourier components $\langle T^K_{00} \rangle$ of the energy density: see figure~\ref{fig:QEfixedK}c,d.  These peaks are not unlike the ones seen at $\Delta\phi \approx 2$ in gold-on-gold collisions more central than $60\%$: see figure~2 of \cite{Adler:2005ee}.  They are quite different from the narrow peak that we find at $80\,{\rm MeV}/c$ (figure~\ref{fig:QEfixedK}a).  At $80\,{\rm MeV}/c$, which is about a quarter the temperature, a hydrodynamic description is probably justified, and we should interpret the narrow peak as a sonic boom.  The analytic form
 \eqn{QEseriesAgain}{
  \langle T^K_{00} \rangle = {\pi^3 T^4 \sqrt{g_{YM}^2 N} \over
   \sqrt{1-v^2}} 
   {3v (1+v^2) \cos\theta \over 2iK}
    {1 \over (1-3v^2\cos^2 \theta)\left( 1 - {ivK \cos\theta \over
      1+v^2} \right) - ivK \cos\theta} + O(K) \,,
 }
which follows from \eno{QtoQ}, \eno{GotTmn}, and \eno{QEseries}, is highly accurate in the infrared limit.  One can see from figure~\ref{fig:QEfixedK} that for $v=0.95$, \eno{QEseriesAgain} loses validity around $K \approx 1$, corresponding to $1\,{\rm GeV}/c$.  In the interesting region of $1$ to a few ${\rm GeV}/c$, $\langle T^K_{00} \rangle$ decreases significantly more quickly with increasing $K$ than the approximation \eno{QEseriesAgain} would indicate.  This falloff may be a positive feature in comparing to data.

Let us enumerate the reasons to treat with caution the connection we allege between our AdS/CFT calculations and the RHIC results on away side jet splitting.
 \begin{enumerate}
  \item $Q^K_E$ is the hardest quantity to compute of the three that we investigated: the system of equations is more formidable, and the nearly singular behavior of $Q^K_E$ near the origin makes numerical evaluations less stable.  Moreover, our analytic approximation \eno{QEseries} is not as precise as for $Q^K_A$ and $Q^K_D$, meaning that we have less extensive checks on numerics.
  \item The $\Delta\phi$ in figure~2 of \cite{Adler:2005ee} is the separation in azimuthal angle, whereas in our figure~\ref{fig:QEfixedK} it is $\pi$ minus the angle between the emission direction and the motion of the heavy quark.
  \item The broad peaks at $\Delta\phi \approx 2$ in \cite{Adler:2005ee} are distinctive only after a subtraction related to elliptic flow.
  \item The peaks of di-jet hadron pair correlations are closer to $\Delta\phi \approx 2$ than to the peak angle $\Delta\phi \approx 2.4$ in figure~\ref{fig:QEfixedK}.
  \item The experimental studies \cite{Wang:2004kf,Adler:2005ee} do not include heavy quark tagging, so most of the away side partons are presumably light quarks or gluons.  But perhaps, for high-angle emission, what matters most is not the quark mass but simply the color current associated with a hard parton.
  \item After a parton leaves the QGP, it fragments, and then its fragmentation products must be detected.  We do not have the expertise to add these important aspects of the physics to our calculations.
  \item The QGP cools, expands, and hadronizes, and its equation of state changes with time as a result.  The conformal result $c_s = 1/\sqrt{3}$ is likely to be a reasonable approximation only in the QGP regime, at temperatures significantly above the deconfinement transition.  As remarked in \cite{Casalderrey-Solana:2004qm}, a steeper emission angle results from a time-averaged speed of sound that could be as low as $c_s \approx 0.33$.  It might be possible to partially mimic the changing equation of state by some deformation of $AdS_5$-Schwarzschild, but it's not clear that the result would have the same status as a first-principles calculation that can be claimed for our analysis.
  \item One of our many idealizations of the true experimental setup is that we replaced the QGP by a thermal medium of infinite extent.  This could mean that we are exaggerating the effects of secondary rescatterings.  It would be desirable to have some position space representations of components of the stress tensor to address this issue.
  \item It remains to us a deep mystery when and why strongly coupled ${\cal N}=4$ gauge theory should be directly compared with real-world QCD\@.  Doing so somewhat above the deconfinement and chiral symmetry breaking transitions is clearly the best hope.  But we return to the basic conundrum: are near-extremal D3-branes merely an analogous system to the QGP, or can they capture the dynamics sufficient precisely to be a useful guide to RHIC physics?
  \end{enumerate}

It is clear from figure~\ref{fig:LobeFig} that as one passes to higher momenta, the peak emission direction becomes more forward, although at the same time the peak keeps broadening.  It would be interesting to compare the dependence of $\langle T_{00}^K \rangle$ on both the magnitude and angle of $\vec{K}$ with two-dimensional histograms of $p_T$ and $\Delta\phi$ for partner hadrons.

As $K$ increases beyond the range shown in figure~\ref{fig:LobeFig}, one eventually passes into the region of high momentum ridges, which become more and more forward as $v \to 1$.\footnote{We thank J.~Casalderrey-Solana for pointing out to us that the peak angle of the high momentum ridges decreases roughly as $1/\gamma$, and for the interesting remark that this behavior may signal some connection with the Landau-Pomeranchuck-Migdal effect (see for example \cite{Vitev:2005yg}).}  As we understand the experimental situation, the away side peak reappears as one increases the momentum window for the hadrons.  An optimistic read of this situation is that AdS/CFT calculations may have some relevance up to an unexpectedly high range of momenta; but perhaps it is more reasonable simply to suppose that sufficiently high-momentum partons can punch through the QGP without much deflection.

There have been other notable theoretical efforts to understand the splitting of the away side jet.  An account of the sonic boom picture can be found in \cite{Casalderrey-Solana:2004qm}.  Investigations of the Cerenkov radiation have been pursued in \cite{Majumder:2005sw,Koch:2005sx}.  The conical flow picture has been seen in \cite{Chaudhuri:2005vc} using the hydrodynamical evolution of QGP, and in \cite{Ruppert:2005uz} using linear response theory.  In \cite{Ruppert:2005uz} the Mach cone picture appears only in the strongly coupled QGP\@.  In comparing with these more phenomenological works, it must be admitted that we have made dramatic and risky idealizations of the experimental setup.  Yet, despite the potential stumbling blocks, it is exciting to see a simple type~IIB string theory construction approaching quantitative comparisons with a data-rich experimental field.

\section*{Acknowledgments}

We thank V.~Gupta and N.~Arora for accommodating our requests for departmental computing resources to run numerical calculations, and we thank D.~Teaney, J.~Casalderrey-Solana, and B.~Zajc for discussions.  This work was supported in part by the Department of Energy under Grant No.\ DE-FG02-91ER40671, and by the Sloan Foundation.  The work of J.F.~was also supported in part by the NSF Graduate Research Fellowship Program.  The work of S.P.~was also supported in part by Princeton University's Round Table Fund for senior thesis research.

\bibliographystyle{ssg}
\bibliography{stress}

\begingroup\raggedright\begin{thebibliography}{10}

\bibitem{Herzog:2006gh}
C.~P. Herzog, A.~Karch, P.~Kovtun, C.~Kozcaz, and L.~G. Yaffe, ``Energy loss of
  a heavy quark moving through N = 4 supersymmetric Yang-Mills plasma,''
  \href{http://xxx.lanl.gov/abs/hep-th/0605158}{{\tt hep-th/0605158}}.

\bibitem{Gubser:2006bz}
S.~S. Gubser, ``Drag force in AdS/CFT,''
  \href{http://xxx.lanl.gov/abs/hep-th/0605182}{{\tt hep-th/0605182}}.

\bibitem{Maldacena:1997re}
J.~M. Maldacena, ``The large N limit of superconformal field theories and
  supergravity,'' {\em Adv. Theor. Math. Phys.} {\bf 2} (1998) 231--252,
  \href{http://xxx.lanl.gov/abs/hep-th/9711200}{{\tt hep-th/9711200}}.

\bibitem{Gubser:1998bc}
S.~S. Gubser, I.~R. Klebanov, and A.~M. Polyakov, ``Gauge theory correlators
  from non-critical string theory,'' {\em Phys. Lett.} {\bf B428} (1998)
  105--114, \href{http://xxx.lanl.gov/abs/hep-th/9802109}{{\tt
  hep-th/9802109}}.

\bibitem{Witten:1998qj}
E.~Witten, ``Anti-de Sitter space and holography,'' {\em Adv. Theor. Math.
  Phys.} {\bf 2} (1998) 253--291,
  \href{http://xxx.lanl.gov/abs/hep-th/9802150}{{\tt hep-th/9802150}}.

\bibitem{Casalderrey-Solana:2006rq}
J.~Casalderrey-Solana and D.~Teaney, ``Heavy quark diffusion in strongly
  coupled N = 4 Yang Mills,''
  \href{http://xxx.lanl.gov/abs/hep-ph/0605199}{{\tt hep-ph/0605199}}.

\bibitem{Sin:2004yx}
S.-J. Sin and I.~Zahed, ``Holography of radiation and jet quenching,'' {\em
  Phys. Lett.} {\bf B608} (2005) 265--273,
  \href{http://xxx.lanl.gov/abs/hep-th/0407215}{{\tt hep-th/0407215}}.

\bibitem{Liu:2006ug}
H.~Liu, K.~Rajagopal, and U.~A. Wiedemann, ``Calculating the jet quenching
  parameter from AdS/CFT,'' \href{http://xxx.lanl.gov/abs/hep-ph/0605178}{{\tt
  hep-ph/0605178}}.

\bibitem{Buchel:2006bv}
A.~Buchel, ``On jet quenching parameters in strongly coupled non- conformal
  gauge theories,'' \href{http://xxx.lanl.gov/abs/hep-th/0605178}{{\tt
  hep-th/0605178}}.

\bibitem{Herzog:2006se}
C.~P. Herzog, ``Energy Loss of Heavy Quarks from Asymptotically AdS
  Geometries,'' \href{http://xxx.lanl.gov/abs/hep-th/0605191}{{\tt
  hep-th/0605191}}.

\bibitem{Caceres:2006dj}
E.~Caceres and A.~Guijosa, ``Drag Force in Charged N=4 SYM Plasma,''
  \href{http://xxx.lanl.gov/abs/hep-th/0605235}{{\tt hep-th/0605235}}.

\bibitem{Friess:2006aw}
J.~J. Friess, S.~S. Gubser, and G.~Michalogiorgakis, ``Dissipation from a heavy
  quark moving through N = 4 super- Yang-Mills plasma,''
  \href{http://xxx.lanl.gov/abs/hep-th/0605292}{{\tt hep-th/0605292}}.

\bibitem{Sin:2006yz}
S.-J. Sin and I.~Zahed, ``Ampere's Law and Energy Loss in AdS/CFT Duality,''
  \href{http://xxx.lanl.gov/abs/hep-ph/0606049}{{\tt hep-ph/0606049}}.

\bibitem{Gao:2006se}
Y.-h. Gao, W.-s. Xu, and D.-f. Zeng, ``Wake of color fileds in charged N = 4
  SYM plasmas,'' \href{http://xxx.lanl.gov/abs/hep-th/0606266}{{\tt
  hep-th/0606266}}.

\bibitem{Armesto:2006zv}
N.~Armesto, J.~D. Edelstein, and J.~Mas, ``Jet quenching at finite 't Hooft
  coupling and chemical potential from AdS/CFT,''
  \href{http://xxx.lanl.gov/abs/hep-ph/0606245}{{\tt hep-ph/0606245}}.

\bibitem{Peeters:2006iu}
K.~Peeters, J.~Sonnenschein, and M.~Zamaklar, ``Holographic melting and related
  properties of mesons in a quark gluon plasma,''
  \href{http://xxx.lanl.gov/abs/hep-th/0606195}{{\tt hep-th/0606195}}.

\bibitem{Avramis:2006ip}
S.~D. Avramis and K.~Sfetsos, ``Supergravity and the jet quenching parameter in
  the presence of R-charge densities,''
  \href{http://xxx.lanl.gov/abs/hep-th/0606190}{{\tt hep-th/0606190}}.

\bibitem{Lin:2006au}
F.-L. Lin and T.~Matsuo, ``Jet quenching parameter in medium with chemical
  potential from AdS/CFT,'' \href{http://xxx.lanl.gov/abs/hep-th/0606136}{{\tt
  hep-th/0606136}}.

\bibitem{Caceres:2006as}
E.~Caceres and A.~Guijosa, ``On drag forces and jet quenching in strongly
  coupled plasmas,'' \href{http://xxx.lanl.gov/abs/hep-th/0606134}{{\tt
  hep-th/0606134}}.

\bibitem{Vazquez-Poritz:2006ba}
J.~F. Vazquez-Poritz, ``Enhancing the jet quenching parameter from marginal
  deformations,'' \href{http://xxx.lanl.gov/abs/hep-th/0605296}{{\tt
  hep-th/0605296}}.

\bibitem{Adler:2002tq}
{\bf STAR} Collaboration, C.~Adler {\em et.~al.}, ``Disappearance of
  back-to-back high p(T) hadron correlations in central Au + Au collisions at
  s(NN)**(1/2) = 200-GeV,'' {\em Phys. Rev. Lett.} {\bf 90} (2003) 082302,
  \href{http://xxx.lanl.gov/abs/nucl-ex/0210033}{{\tt nucl-ex/0210033}}.

\bibitem{Wang:2004kf}
{\bf STAR} Collaboration, F.~Wang, ``Measurement of jet modification at RHIC,''
  {\em J. Phys.} {\bf G30} (2004) S1299--S1304,
  \href{http://xxx.lanl.gov/abs/nucl-ex/0404010}{{\tt nucl-ex/0404010}}.

\bibitem{Adler:2005ee}
{\bf PHENIX} Collaboration, S.~S. Adler {\em et.~al.}, ``Modifications to
  di-jet hadron pair correlations in Au + Au collisions at s(NN)**(1/2) =
  200-GeV,'' \href{http://xxx.lanl.gov/abs/nucl-ex/0507004}{{\tt
  nucl-ex/0507004}}.

\bibitem{Balasubramanian:1999re}
V.~Balasubramanian and P.~Kraus, ``A stress tensor for anti-de Sitter
  gravity,'' {\em Commun. Math. Phys.} {\bf 208} (1999) 413--428,
  \href{http://xxx.lanl.gov/abs/hep-th/9902121}{{\tt hep-th/9902121}}.

\bibitem{Gubser:1996de}
S.~S. Gubser, I.~R. Klebanov, and A.~W. Peet, ``Entropy and Temperature of
  Black 3-Branes,'' {\em Phys. Rev.} {\bf D54} (1996) 3915--3919,
  \href{http://xxx.lanl.gov/abs/hep-th/9602135}{{\tt hep-th/9602135}}.

\bibitem{Teaney:2006nc}
D.~Teaney, ``Finite temperature spectral densities of momentum and R- charge
  correlators in N = 4 Yang Mills theory,''
  \href{http://xxx.lanl.gov/abs/hep-ph/0602044}{{\tt hep-ph/0602044}}.

\bibitem{Danielsson:1998wt}
U.~H. Danielsson, E.~Keski-Vakkuri, and M.~Kruczenski, ``Vacua, propagators,
  and holographic probes in AdS/CFT,'' {\em JHEP} {\bf 01} (1999) 002,
  \href{http://xxx.lanl.gov/abs/hep-th/9812007}{{\tt hep-th/9812007}}.

\bibitem{QXwebsite}
http://wwwphy.princeton.edu/$\sim$ssgubser/papers/stress/index.html.

\bibitem{JacakPrivate}
B. Jacak, private communication.

\bibitem{Casalderrey-Solana:2004qm}
J.~Casalderrey-Solana, E.~V. Shuryak, and D.~Teaney, ``Conical flow induced by
  quenched QCD jets,'' {\em J. Phys. Conf. Ser.} {\bf 27} (2005) 22--31,
  \href{http://xxx.lanl.gov/abs/hep-ph/0411315}{{\tt hep-ph/0411315}}.

\bibitem{Vitev:2005yg}
I.~Vitev, ``Large angle hadron correlations from medium-induced gluon
  radiation,'' {\em Phys. Lett.} {\bf B630} (2005) 78--84,
  \href{http://xxx.lanl.gov/abs/hep-ph/0501255}{{\tt hep-ph/0501255}}.

\bibitem{Majumder:2005sw}
A.~Majumder and X.-N. Wang, ``LPM interference and Cherenkov-like gluon
  bremsstrahlung in dense matter,'' {\em Phys. Rev.} {\bf C73} (2006) 051901,
  \href{http://xxx.lanl.gov/abs/nucl-th/0507062}{{\tt nucl-th/0507062}}.

\bibitem{Koch:2005sx}
V.~Koch, A.~Majumder, and X.-N. Wang, ``Cherenkov radiation from jets in
  heavy-ion collisions,'' {\em Phys. Rev. Lett.} {\bf 96} (2006) 172302,
  \href{http://xxx.lanl.gov/abs/nucl-th/0507063}{{\tt nucl-th/0507063}}.

\bibitem{Chaudhuri:2005vc}
A.~K. Chaudhuri and U.~Heinz, ``Effect of jet quenching on the hydrodynamical
  evolution of QGP,'' \href{http://xxx.lanl.gov/abs/nucl-th/0503028}{{\tt
  nucl-th/0503028}}.

\bibitem{Ruppert:2005uz}
J.~Ruppert and B.~Muller, ``Waking the colored plasma,'' {\em Phys. Lett.} {\bf
  B618} (2005) 123--130, \href{http://xxx.lanl.gov/abs/hep-ph/0503158}{{\tt
  hep-ph/0503158}}.

\end{thebibliography}\endgroup

\end{document}